%% file: main.tex
\documentclass[10pt, conference, letterpaper]{IEEEtran}
\IEEEoverridecommandlockouts
\usepackage{cite}
\usepackage{amsmath,amssymb,amsfonts}
\usepackage{algorithmic}
\usepackage{graphicx}
\usepackage{textcomp}
\usepackage{xcolor}
\usepackage[english]{babel}
\usepackage[utf8x]{inputenc}
\usepackage{amsmath}
\graphicspath{/Figures}
\usepackage{epstopdf}
\usepackage[linesnumbered,ruled]{algorithm2e}
\usepackage{array}
\usepackage{caption}
\usepackage{subcaption}
\usepackage{float}
\usepackage{todonotes}
\usepackage{paralist}

\def\BibTeX{{\rm B\kern-.05em{\sc i\kern-.025em b}\kern-.08em
    T\kern-.1667em\lower.7ex\hbox{E}\kern-.125emX}}
    
\begin{document}

\title{A Network-centric Framework for\\ Auditing Recommendation Systems}

\author{
\IEEEauthorblockN{Abhisek Dash, Animesh Mukherjee, Saptarshi Ghosh}
\IEEEauthorblockA{Department of Computer Science and Engineering \\
Indian Institute of Technology Kharagpur, India\\
dash.abhi93@iitkgp.ac.in, animeshm@cse.iitkgp.ac.in, saptarshi@cse.iitkgp.ac.in}
}

\maketitle

\begin{abstract}
To improve the experience of consumers, all social media, commerce and entertainment sites deploy Recommendation Systems (RSs) that aim to help users locate interesting content. These RSs are black-boxes -- the way a chunk of information is filtered out and served to a user from a large information base is mostly opaque. 
No one except the parent company generally has access to the entire information required for auditing these systems -- neither the details of the algorithm nor the user-item interactions are ever made publicly available for third-party auditors. Hence auditing RSs remains an important challenge, especially with the recent concerns about how RSs are affecting the views of the society at large with new technical jargons like ``echo chambers'', ``confirmation biases'', ``filter bubbles'' etc. in place. 
Many prior works have evaluated different properties of RSs such as diversity, novelty, etc. However, most of these have focused on evaluating static snapshots of RSs. Today, auditors are not only interested in these static evaluations on a snapshot of the system, but also interested in how these systems are affecting the society in course of time. In this work, we propose a novel network-centric framework which is not only able to quantify various static properties of RSs, but also is able to quantify dynamic properties such as how likely RSs are to lead to polarization or segregation of information among their users. We apply the framework to several popular movie RSs to demonstrate its utility.
\end{abstract}


\input{Introduction}
\input{Related}
\input{Framework}

\input{Diversity}

\input{Information_Segregation}
\input{Conclusion}

\bibliographystyle{IEEEtran}
\bibliography{sample-bibliography}


\end{document}

%% file: Introduction.tex
\section{Introduction}

\noindent
The digital abode is full of choices. Users buy products, book trips, pay bills and watch movies online, and in all these scenarios they are presented with multiple choices. The subsequent decisions are known to be influenced by the choice environment, i.e., \textit{how} the choices are presented to the users. 
People, having bounded rationality, are mainly driven by a set of heuristics and/or inherent biases for making a decision. Hence, the choice architects can use these heuristics and biases to manipulate the choice environment to guide the users' actions by gently ``nudging'' them toward certain choices. 
Such ``presentation biases'' (nudges) are very relevant in the context of different information filtering systems~\cite{schneider2018digital, baeza2018bias}.
Especially, Recommender Systems (RSs) have evolved as an inescapable module of any online platform ranging from social networking to e-commerce to entertainment sites.
RSs play an instrumental role in deciding the profit margins of almost all e-companies.
From the client side, an RS helps users find relevant and novel items from the enormous information base, which saves their time and fulfills their interests. 

Given all the positive outcomes of RSs and the important role they are playing in filtering information out to the society, the intelligence of these systems needs to be monitored periodically. 
``Algorithm auditing'' provides researchers, designers, and users new ways to understand the algorithms that increasingly shape our online life and opinion, and diagnose the unwanted consequences of algorithmic systems. 
For a precise audit of RSs, the knowledge about the algorithm itself, the entire information/product base on which the algorithm works, and the user's actions with the recommendations produced are the major prerequisites for an auditor. However, all these details are never publicized by the commercial websites, which lead to the research question that we put forward in this paper --
\textbf{RQ}: How does a third party audit a Recommendation System without having all these subtle details? 

To answer the above question, in this paper, we present a novel network-based technique that enables us to extract important parameters for auditing RSs. 
In particular, we consider two important properties -- (i)~diversity of the recommendations provided by the RSs, 
and (ii)~the extent of information segregation/polarization that the RSs induces among their user population. 
At this point, we would like to clarify that standard properties like relevance, accuracy etc. of RSs can also be evaluated by our proposed framework; however, we believe that these do not qualify as the most important auditing parameters, since there have been a lot of complaints in the literature citing how too much concentration on improving relevance has led to the unwanted fracture of the `global village' of information into `tribes'~\cite{hosanagar2013will}. 
Pariser coined the term \textit{filter bubble} to succinctly express this `worry' -- a phenomena resulting in a self-reinforced pattern of narrowing exposure~\cite{pariser2011filter}.
 
Our framework constitutes of a directed weighted network where the nodes correspond to different items present on a commercial website, and there is a directed edge from an item $i$ to an item $j$ if item $j$ features in the recommendation list that is shown on the page of the item $i$. We refer to this network as the {\it recommendation network.}\footnote{Note that the network is independent of the user information, 
and does not assume knowledge of user-item interactions.}
The different attributes of the items (e.g., genre or type of movies, 
topics of news articles, etc.)
can be encoded as node properties. Note that the framework is different from the standard item-item similarity network, which essentially is a complete undirected graph and does not depend on recommendation outputs.
At the same time this recommendation network has the ability to model the nudges that users, exposed to these digital environments, could experience.

\vspace{2mm}

\noindent\textbf{Contributions}: We choose the movie domain for the purpose of our experiments.
In particular, we apply our framework on three popular movie recommendation sites -- IMDb (\url{www.imdb.com}), Google Play (\url{play.google.com/store/movies}), and Netflix (\url{dvd.netflix.com}). 
The motivation behind investigating the movie domain is driven by the recent claims of bias in the Recommendation Systems deployed at online movie sites such as Netflix\footnote{Netflix’s Recommendation Algorithm Is Borderline Racist: \url{https://nylon.com/articles/netflix-race-algorithm}}, which makes it important to audit movie recommendation systems. 

Based on the formulation of a recommendation network, we make the following contributions in this paper.
\begin{compactitem}
	\item We put forward novel quantifications of diversity. The first among these is based on {\it mixing patterns} of the recommendation network~\cite{newman2003structure}, where the movies are binned based on their genre.
	We observe that IMDb and Netflix present higher diversity than Google Play. Some of the interesting findings are that for Google Play, the movies in $7$ genres - `comedy', `animation', `action and adventure', `Indian cinema', `horror' and `documentary' - almost always recommend movies of the same genre itself. 
	For IMDb, barring the two genres `adult' and `reality TV', the recommendations from all other genres are quite diverse and take the user to many other genres. Similar is the case for Netflix. 
	
	\item The second set of measures is based on graph-based notions of popularity like in-degree and PageRank. If one bins the items based on in-degree or PageRank, one observes an universal phenomena across all the three websites -- the top bins have very diverse set of recommendations, often leading to the middle and the bottom bins; in contrast, the items in the bottom bin (i.e., the non-popular items in the long tail) mostly recommend other items from  the bottom bin itself. 
	This observation though is not surprising because any recommendation system is supposed to push the users to browsing the `long tail', i.e., items from the bottom bin. 
	Further, the top bin of IMDb seem to be more diverse than the other two platforms. Although the top bin of Netflix recommends higher fraction of movies from the bottom bin than IMDb, it recommends very low number of movies from the middle bin. For Google Play the number of recommendations going from the top to other bins is very low compared to the other two platforms. 
	
	\item As a following step, we model diversity as a process, to analyze the experience of a user who uses the RSs over a long period of time. 
	We simulate a user as a random surfer~\cite{page1999pagerank} walking on the recommendation network. 
	The randomness of the surfer corresponds to the propensity of a user to follow recommendations of the system, and is controlled by the teleportation probability $t_p$. A surfer with $t_p = 0.0$ would always follow recommendations given by the RSs, i.e., traverse the network from one node to the other following the edges in the network.
	Whereas, a surfer with $t_p = 1.0$ would jump from one node to another randomly, agnostic to the edges, i.e., without following recommendations. 
	We measure the diversity experienced by a user by the entropy of the distribution of movies (across the genres) that the user observes during the walk.
	An interesting observation is that the diversity in the recommendations received by a surfer increases rapidly as the propensity to follow recommendations decreases (i.e., with increasing teleportation probability) till a point after which it flattens. 
    
\if{0}
\item To understand how the view of the surfer might get affected due to diversity, we compare the distribution of movies across the genres generated due to the random walk with the global distribution of all movies across these genres. 
	In particular, we compute the K-L divergence of the simulated distribution from the global distribution. We make three key observations here -- 
	(i)~systems which are more diverse like IMDb and Netflix present a view to the surfer that is far from the global distribution, as compared to a less diverse system like Google Play, 
	(ii)~increasing teleportation probability brings the view significantly closer to the global distribution for less diverse systems like Google Play, and 
	(iii)~larger walk lengths brings the surfer closer to the global distribution for all the systems.
\fi
    
\item 
Apart from diversity, another important question in auditing RSs is to what extent the RSs is leading to information segregation or polarization among the user population, leading different parts of the population to form different opinions.
Opinion formation often is led by the type of content an individual or a group of users is exposed to. Hence, polarization can also be thought of as a process~\cite{dandekar2013biased}. 
We study the extent of segregation induced by the three movie RSs among a population of users. 
To this end, we simulate the random walk discussed above for as many as 1100 different users organized in 110 groups, each group having a certain propensity to follow the recommendations (one group corresponding to a particular value of $t_p \in [0,1]$) and a certain start point on the network. 
Using the final set of movies that each user hopped (read or viewed) during the random walk, we compute two measures for information segregation -- \textit{concentration} and \textit{evenness} (defined in~\cite{chakraborty2017quantifying}) -- to investigate the extent of segregation in the three RSs. 
The key observations from this analysis are that (i)~following the IMDb recommendations, a user will have a considerably less polarizing / segregating experience of different movie genres, as compared to that for Netflix and Google Play, and 
(ii)~among Netflix and Google Play, the latter induces more segregation among the users.

\end{compactitem}

\noindent
We believe that our prime contribution lies in representing RSs as networks, to not only study the diversity of RSs from various angles but also to add significant explanatory power through the dynamic measures that can be used to examine the process of information segregation induced by the RSs.

%% file: Related.tex
\section{Background and Related Work}

RSs are often the primary view through which a user has access to a large information base that is otherwise difficult to navigate.
Though the recommendations provided by RSs might satisfy the immediate needs of a user, on a long term the user might be stifled into an unchanging environment. 
Such an unchanging environment was given the name of \textit{filter bubble} by Eli Pariser~\cite{pariser2011filter}. 
Following these studies, there has been lot of research on diversity and explainability of recommendations and dynamics of polarization.
We discuss few of these studies in this section.

\noindent{\bf Diversity and novelty in RSs}: In the context of RSs, novelty and diversity are different though related notions. 
The novelty of an item (e.g., movie) generally refers to how different the item is from what a user has already observed till the point of recommendation, while diversity generally is defined over a set of such items (that are recommended together). 
It is generally agreed that the primary challenge in improving diversity and novelty is their trade-off with the accuracy/relevance of recommendations.

Several studies have proposed ways to measure diversity and novelty.
Nguyen et. al.~\cite{nguyen2014exploring} examined the longitudinal impact of collaborative filtering RSs on users, and
evaluated diversity based on information encoded in user-generated tags.
Zhou et. al.~\cite{zhou2010solving} explained novelty as the mean self-information of the items recommended, which is evaluated as the inverse user frequency.
Santini and Castells~\cite{santini2011evaluation} developed measures that evaluated novelty and diversity with a fuzzy interpretation.
Vargas and Castells studied formal characterizations for evaluation of novelty and diversity from an end user's perspective~\cite{vargas2011rank} . They derived few metrics considering the item position and relevance of the items in the recommended list, which is generally not taken into consideration while evaluating diversity. 
Lathia et. al.~\cite{lathia2010temporal} brought {\it time} into consideration for how the recommendations change in course of time, i.e., how the RSs reacts to or evolves with time and at different depths of recommendation.

As is evident from the discussion above, several different notions of diversity (and related concepts like novelty) have been proposed, which can broadly be divided into two categories --
(i)~the measures which assume that data about user-item interactions is available (e.g., user-clicks or user-ratings of different items), and
(ii)~measures which do not rely on user-item interaction data~\cite{ricci2011introduction}. 
Also, evaluation of diversity can be done in local and global granularities. 
Analysis done specifically on the local recommendation lists at the site of each item is referred to as local analysis~\cite{ziegler2005improving, ricci2011introduction,bradley2001improving}; on the other hand, analysis done over the entire universe of items is referred to as global analysis~\cite{ricci2011introduction,szlavik2011diversity,adomavicius2011maximizing}. 

\if{0}

%
%

We now describe various measures for diversity of RSs, grouped according to
(i)~local and global measures, and (ii)~whether they assume user-item interaction information (as explained above). 

\subsubsection{Global diversity measures with user-item information}

\begin{itemize}
	
	\item \textbf{Global long-tail novelty:}~\cite{ricci2011introduction}
	Inspired by the analogy of \textit{Inverse Document Frequency (IDF)} in Information Retrieval literature, an \textit{Inverse User Frequency (IUF)} has been proposed. 
	Based on the observed user-item interaction, this can be estimated as $IUF = -\log_{2}{\frac{|U_i|}{|U|}}$, where $U_{i}$ denotes the set of all users who have interacted with item $i$ and $U$ is the set of all users. Thus the novelty of a recommendation list can be quantified as the mean IUF of the list of recommended items.
%
	\item \textbf{Aggregate diversity:} is a measure to evaluate the extent to which the item universe has been exposed to different users. This notion can be formalized in many different ways, e.g. Gini coefficient~\cite{fleder2009blockbuster} or Shannon entropy~ \cite{szlavik2011diversity}.

	\item {\bf Network-based diversity measures:} Many studies considered a user-item bipartite network, where nodes in the two partite-sets represent users and items respectively, and the edge $(u, i)$ implies that user $u$ has interacted with item $i$. E.g., \cite{adomavicius2011maximizing} used this approach for maximizing recommendation diversity based on maximum flow or maximum bipartite matching computations. 
\end{itemize}

\subsubsection{Local diversity measures with user-item information}
\begin{itemize}
	
	\item \textbf{User-specific unexpectedness:} \cite{murakami2007metrics} of a recommended list $R_u$ given to a user $u$ is defined as the average dissimilarity between the recommended items ($R_u$) and all the items that $u$ has come across till the point of recommendation ($J_u$).  
	
\end{itemize}

\subsubsection{Global diversity measures without user-item information}

\begin{itemize}
	\item \textbf{Long-tail novelty:} The global novelty of an item $i$ can be described as the inverse of popularity of the item: 
	$Novelty(i) = \log (\frac{1}{Popularity(i)})$
	Here popularity of an item can be measured as how often an item is recommended on the page of different items.
	
	\item {\bf Network-based diversity measures:} \cite{celma2008new} 
	constructed a graph where all the items are nodes and the weighted edges define the similarity among the items. They performed both user-centric and item-centric analyses to detect whether network topology plays any part in diversity and novelty of recommendations. Rather than assessing novelty just in terms of the long-tail items that are directly recommended, they analyzed the paths leading to these novel items. 
	
\end{itemize}

\subsubsection{Local diversity measures without user-item information}

\begin{itemize}
	\item \textbf{Average intra-list diversity:}
	This is one of the most frequently used diversity metric and was first proposed by \cite{ziegler2005improving} where it was denoted as Intra-List Similarity and elaborated as a decreasing function. Mathematically, intra-list diversity of a set of recommended items is defined as the mean pairwise diversity of the items in the list.
	Other studies such as \cite{smyth2001similarity} have also used this measure.
	
	\item\textbf{Source list diversity:} refers to the average of how dissimilar the recommended items $j$ in recommended list $R_i$ are from the `source item' $i$ for which they are recommended. 
	
	\item\textbf{Average unexpectedness:}~\cite{ricci2011introduction} Unexpectedness of an item $j$ in the recommendation list $R_i$ is defined as the product of the novelty of the item and how different is the given item from the source item $i$.
\end{itemize}
\fi

\noindent {\bf Dynamics of polarization}: Many empirical studies show that homophily, i.e., greater interaction among like-minded individuals often lead to polarization~\cite{gilbert2009blogs}.
Polarization can be thought of as a measure of the ideological state of the population in a society. 
With the advent of Internet, the increased diversity of information sources coupled with the tailoring mechanisms like personalization has an echo chamber effect that can result in increased polarization. 
A majority of works attempt to explain polarization through variants of a well-known mathematical model for opinion formation proposed by DeGroot~\cite{degroot1974reaching}. 
Dandekar et al.~\cite{dandekar2013biased} analyzed the polarizing effects of three recommendation algorithms -- SimpleSALSA, SimplePPR and SimpleICF -- over a natural model of the underlying user-item graph.

Two competing theories of opinion polarization have been proposed in earlier works. One school of thought assumes that opinions are strengthened when like-minded individuals interact~\cite{dandekar2013biased}; the other school claims that exposure to differing views and their subsequent rejections lead to polarization~\cite{baldassarri2008partisans}. 
Hence, polarization is not a property of a state of the society, rather it is a property of the dynamics through which individuals form opinions. 

Opinion formation dynamics can be thought to be polarized if they result in an increased divergence of opinions or access to widely different pieces of information based on the group an individual belongs to. Following the line of work of Massey and Denton~\cite{massey1993american} in residential segregation, Chakraborty et. al.~\cite{chakraborty2017quantifying} present a notion of information segregation by considering bipartite matching between different groups and information units they have access to. We adopt some of these measures to study the extent of segregation in RSs. 

\noindent {\bf Novelty of present work}: We present a novel network-based framework for effective auditing of RSs. The network is built only from recommendation outputs; hence, unlike the traditional user-item networks, no knowledge of user-item interactions is assumed. 
The recommendation network (directed, based on recommendations) proposed in this work is also completely different from item-item similarity networks used in prior works (undirected complete graph agnostic to actual recommendations).
Additionally, most prior measures consider a static snapshot of the RSs; whereas, the proposed framework gives a way to model the interactions of a user with the RSs over a period of time.
Using this framework we propose various novel techniques to measure the diversity and segregation of RSs.

\if{0}
It is clear from the discussion above that several different measures for diversity in a RSs have been proposed. 
The different measures capture different notions of diversity, and it is not clear how to interpret the values produced by the different measures. 
Importantly, the measures which rely on user-item interaction data cannot be applied by third party auditing of RSs, since such data is seldom made public by commercial RSs. 

The framework proposed in this work encompasses all the diversity measures which do not assume user-item information. Hence, applying this framework will 
make it easier to compare between different measures.
Additionally, most of the measures discussed above provide a static measure of diversity. On the other hand, we propose a way to simulate the long-term interaction of a user with the RSs, and to quantify how the experience of the user varies with time. 
In summary, the proposed framework looks into two important aspects, i.e., how diverse an underlying recommendation system is and how polarizing can it be with due course of time.
\fi

%% file: Framework.tex
\section{Framework for auditing RSs}\label{framework}

This section describes the proposed network-based framework for auditing RSs. 
We also describe the datasets that we use to demonstrate the utility of the framework.

\subsection{Network construction}

We propose to model the {\it output} (recommendations) of RSs by directed networks, where each node is an item (out of the universe of items to be recommended) and the directed edge $i \rightarrow j$ implies that the item $j$ is included in the recommendation list shown on the page of item $i$. 
We denote such a network as a `Recommendation Network (RN)'.

While constructing the RN, we do {\it not} assume the availability of user-item interaction data, since such data is generally not publicly available. 
We only assume the data of which items are recommended on the page of a certain item; 
this data can easily be obtained from the RSs website even by third-party auditors.

The edges $i \rightarrow j$ in a RN can be unweighted or weighted based on some similarity measure $sim(i,j)$ 
between the items.
Alternatively, the edges  $i \rightarrow j$ can also be weighted based on the {\it rank} at which $j$ is shown on the page of $i$.

Note that, while some prior works have adopted network-based measures for RSs (e.g., user-item bipartite networks and item-item similarity networks, as surveyed in the Related Work section), our recommendation network is fundamentally different from the networks in the prior works. 
Though the RN has some similarity to item-item similarity networks (both have items as nodes, and edges can be weighted based on item similarity), the construction of the two networks is very different. 
The construction of RN is entirely focused on the recommendation outputs, while, item similarity networks consider the similarity among all products in the product space. Hence, item similarity networks are theoretically undirected complete graphs, while the RN is a directed graph.

\subsection{User modeling on the proposed RN}\label{sub:Umodel}

For a third-party auditor, the unavailability of user-logs is one of the most challenging drawbacks while auditing the RSs. 
To circumvent this problem, we attempt to model the process of users browsing the recommendations as a random walk over the RN. 

We assume that a user will start with viewing a particular item, 
and then choose one of the items recommended on the page of the viewed item. Alternatively, the user can randomly choose some item from among the universe of items.
We simulate such a user as a random surfer~\cite{page1999pagerank} who performs a walk (random / biased) over the RN. 
Different users can have different preferences about whether to follow the recommendations, or whether to select the next item by herself. We model the propensity of a user to follow the recommendations as the {\it teleportation probability} $t_p$ of the walk which varies in $[0,1]$~\cite{page1999pagerank}.
A user having $t_p = 0.0$ always follows the recommendations, i.e., chooses the next item from the list of items recommended on the page of the last viewed item.
On the other extreme, a user having $t_p = 1.0$ never follows the recommendations, and randomly chooses the next item to view.
We experiment with users having $t_p = 0.0, 0.1, 0.2, \ldots 0.9, 1.0$ to cover all types of users.
A special case of a surfer having $t_p = 0.0$ (who always follows recommendations) is a {\it non-stochastic} surfer, who always selects the top-ranked recommendation in the ranked recommended list shown on the page of the last viewed item. 
We denote this non-stochastic surfer as $t_p = 0.0^{*}$. 

As a user interacts with the RSs (i.e., performs a walk over the RN), he/she accesses a certain set of items of different types (e.g., views a set of movies from different genres). 
The distribution of the different types / genres seen by the user, in the course of a walk, is henceforth referred to as the {\bf``observed distribution'' of the user}, which will be used to quantify the diversity of information that the user is exposed to. 

\subsection{Datasets for applying the framework}
\label{sub:datasets}

While the proposed framework can be applied to RSs in any domain (e.g., e-commerce, news media, social friendship recommendations), for the present work, we chose to apply it to {\it movie recommendation systems}. As outlined in the introduction, the choice is motivated by various factors. 
First, there are increasing concerns about various forms of bias in movie recommendation systems, which motivated us to audit movie RSs.
Also, different movie recommendation sites index the same universe of movies, thus making the comparison meaningful (whereas, in domains like online news, the news articles will differ widely between different sites).

We choose three movie recommendation platforms -- (i)~IMDb (\url{www.imdb.com}), (ii)~Google Play (\url{play.google.com/store/movies}), and (iii)~Netflix DVD rental service (\url{dvd.netflix.com}) -- for the present study. Note that the last two are online service providers while the first one is an online database. 
These choices allow us to investigate how the two media service providers compare in diversity and information segregation compared to the database.  

Each of these websites show a ranked list of recommendations on the page of every movie.
We designed snowball sampling (BFS) crawlers for each of these websites. We seeded the crawler with an initial movie, crawled all recommendations shown on the page of the seed movie, and pushed the recommendations to a queue, and repeated the process on the items in the queue. We continued the crawls till the queue was exhausted, to ensure that we collect the whole universe of items. 
The total number of movies whose data we could collect from the three sites are shown in Table~\ref{netStat} (number of nodes).

\begin{table}[tb]
	\noindent
	\small
	\centering
	\begin{tabular}{ |p{1.5cm}|r|r|p{1cm}|p{1.85cm}| }
		\hline
		RN & Nodes & Edges & Avg. Degree & Reciprocating Edges  \\
		\hline
		IMDb & 172,582 & 1,463,966 & 8.483 &  311,424 (21.27\%)\\
		\hline
		Google Play & 2,143 & 40,663 & 18.975 & 12,107 (29.77\%)\\
		\hline
		Netflix & 24,016 & 166,232 & 7.316 & 9,456 (5.66\%)\\
		\hline
	\end{tabular}	
	\caption{{\bf Statistics of Recommendation Networks of three popular movie RSs -- IMDb, Google Play, and Netflix.}}
	\label{netStat}
	\vspace*{-7mm}
\end{table}

We took some precautions to ensure that the comparative analysis of the various RSs is meaningful.
It is possible for websites to sense the location from which a view is being made, and to tailor/personalize the view toward that location. To account for the effects of such personalization, we perform all the crawls from the same IP address. Additionally, we ensure that the crawls are done without logging in, and without any session history being stored. We also ensure that the data from all three sites are gathered over the same time duration (of about two weeks). 

\vspace{1mm}
\noindent {\bf Genre of the movies:} Along with the recommendations, we also collected meta-data of the movies. While IMDb stores extensive meta-data (e.g., actors, genre, directors, screen-time, etc.), other sites do not store as much meta-data about the movies. 
One attribute that is stored across all sites is the {\it genre} of the movies. 
However, the genres in different sites are different. Google Play specifies 15 different genres, some of which are `action and adventure', `mystery and suspense', etc.
IMDb and Netflix both have 29 different genres, some of which are `action', `romance', `crime', `comedy', `animation', `sci-fi', etc. 
Note that the same movie can have multiple genres, e.g., the movie `Titanic' has genres `drama' and `romance', while `The Godfather' has genres `crime' and `drama' on IMDb.

\subsection{Recommendation networks for movie RSs}

We create the RN for the three movie RSs, and report some basic statistics in Table~\ref{netStat}. IMDb has the largest set of movies, followed by Netflix and Google Play. 
However, Google Play has the highest average node-degree. 

\if{0}
The degree distributions of the RNs (both in-degree and out-degree distributions) are shown in Figure~\ref{DD}. In all the three RNs, the out-degree distribution peaks at a certain value because the number of recommendations shown on the webpage of a movie is usually constant. The in-degree distributions show a decreasing curve with in-degree; specifically the in-degree distribution for Netflix resembles a power-law behavior.

\begin{figure*}[tb]
	\centering
	\begin{subfigure}{0.3\textwidth}
		\centering
		\includegraphics[width=\textwidth, height=3.5cm]{Figures/Deg_IMDb.png}
		\caption{IMDb}
		\label{Deg1}
	\end{subfigure}%
	\begin{subfigure}{.30\textwidth}
		\centering
		\includegraphics[width=\textwidth, height=3.5cm]{Figures/Deg_GP.png}
		\caption{Google Play}
		\label{Deg2}
	\end{subfigure}
	\begin{subfigure}{.30\textwidth}
		\centering
		\includegraphics[width=\textwidth, height=3.5cm]{Figures/Deg_Netflix.png}
		\caption{Netflix}
		\label{Deg3}
	\end{subfigure}
	\vspace*{-2mm}
	\caption{{\bf In-degree and out-degree Distributions of the three RNs obtained from IMDb, Google Play and Netflix. The out-degree distributions peak at a certain point since a constant number of recommendations are shown on all pages. The in-degree distributions fall quickly with increasing node-degree.}}
	\label{DD}
	\vspace*{-4mm}
\end{figure*}

We note that there are significant differences among the RNs, even in the neighborhood of the same movie. To demonstrate this difference, Figure~\ref{Ego} shows the 1.5 degree ego-centric network of a particular node (the movie `The Godfather') that is present in all the three RNs.

\begin{figure*}[tb]
	\centering
	\begin{subfigure}{.30\textwidth}
		\centering
		\includegraphics[width= 5cm, height=4.5cm]{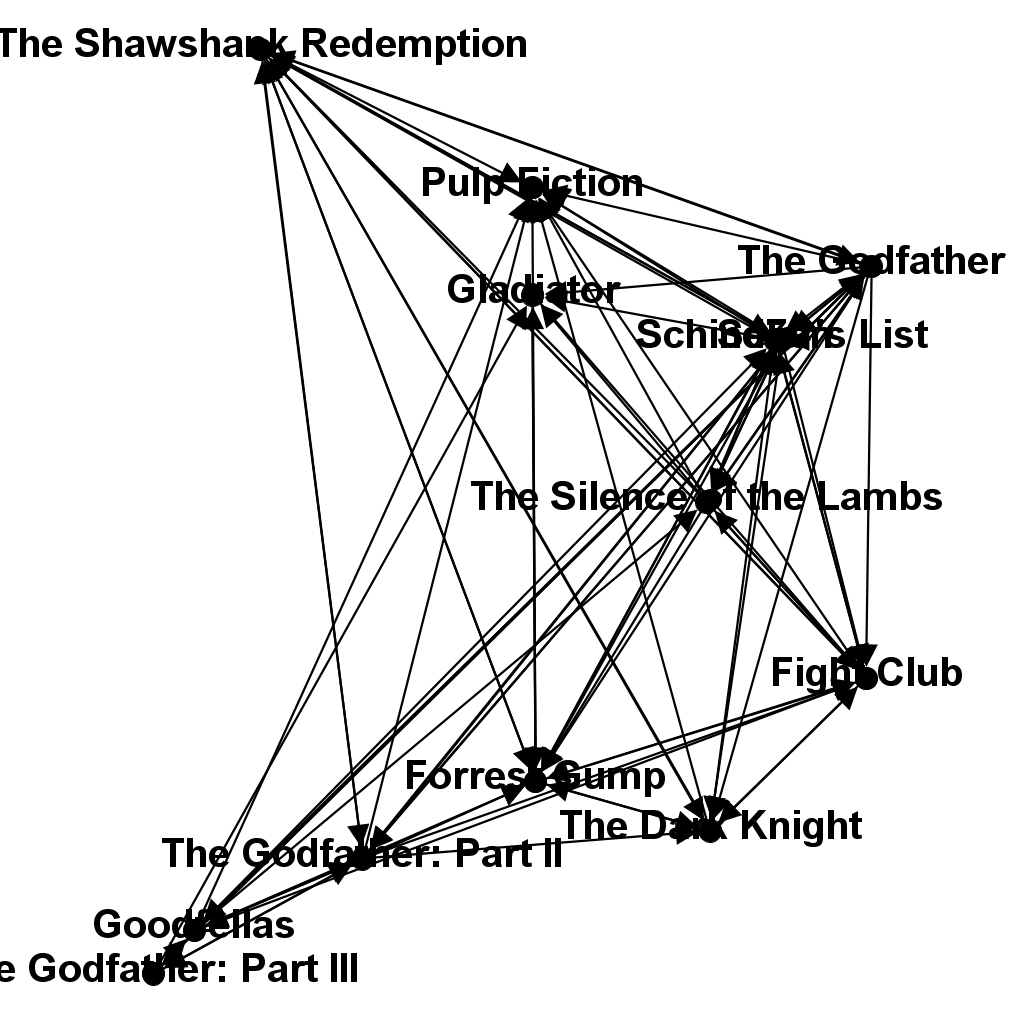}
		\caption{IMDb RN}
		\label{Ego1}
	\end{subfigure}%
	\begin{subfigure}{.30\textwidth}
		\centering
		\includegraphics[width= 5cm, height=4.5cm]{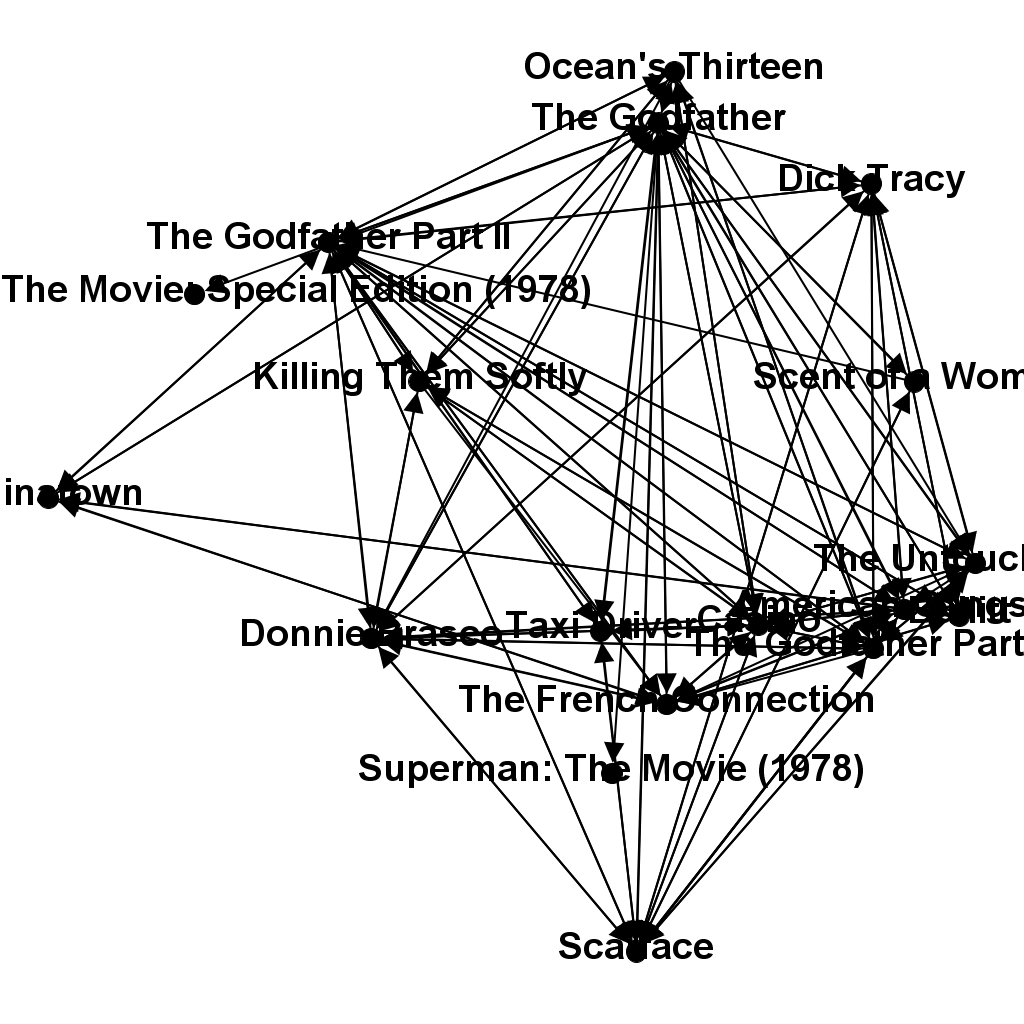}
		\caption{GoogleFPlay RN}
		\label{Ego2}
	\end{subfigure}
	\begin{subfigure}{.30\textwidth}
		\centering
		\includegraphics[width= 5cm, height=4.5cm]{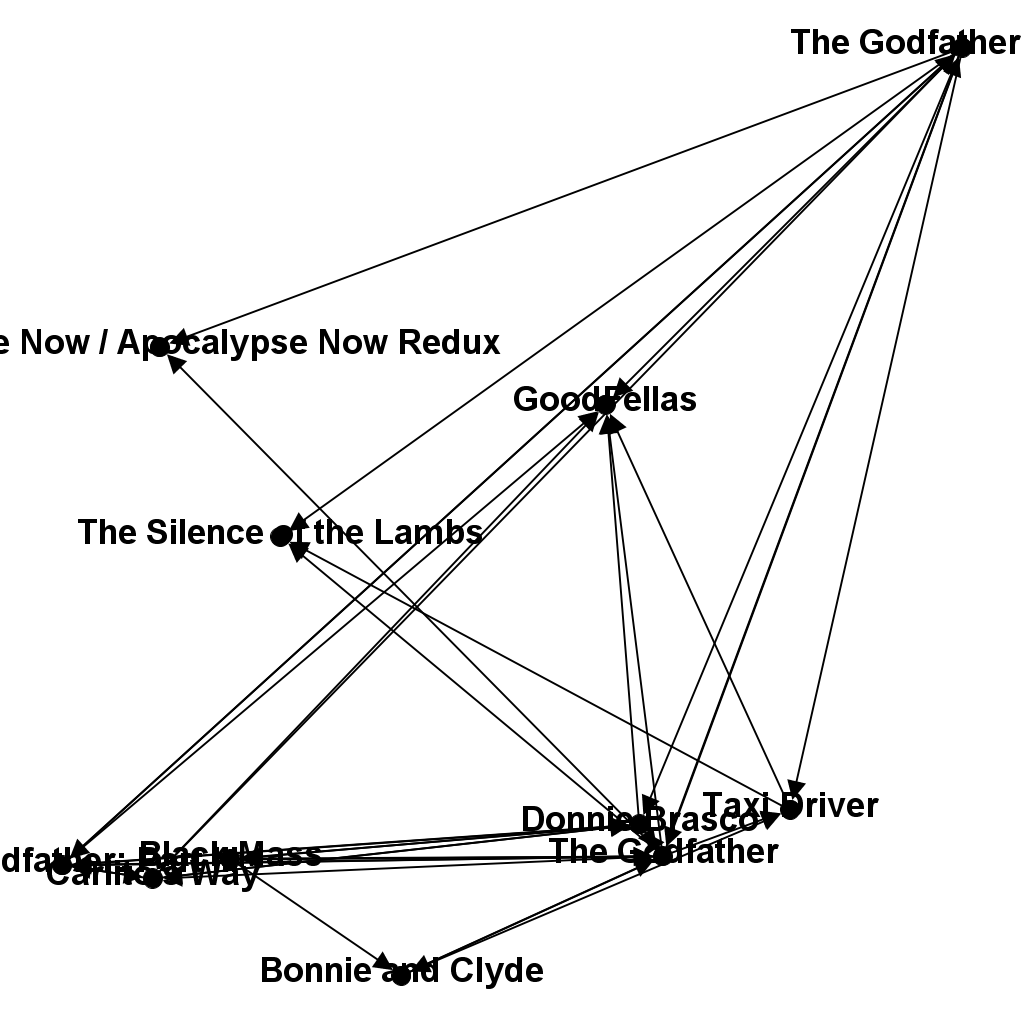}
		\caption{Netflix RN}
		\label{Ego3}
	\end{subfigure}
	\vspace*{-2mm}
	\caption{{\bf Ego-centric network for the movie ``The Godfather'' from all the three recommendation network. We have considered the 1.5 degree ego-centric network of the mentioned movie i.e. all its out-neighbors and their neighbors.}}
	\label{Ego}
	\vspace*{-5mm}
\end{figure*}
\fi

\begin{figure}[tb]
\vspace*{-5mm}
	\centering
	\begin{subfigure}{.48\columnwidth}
		\centering
		\includegraphics[width=\textwidth, height=3cm]{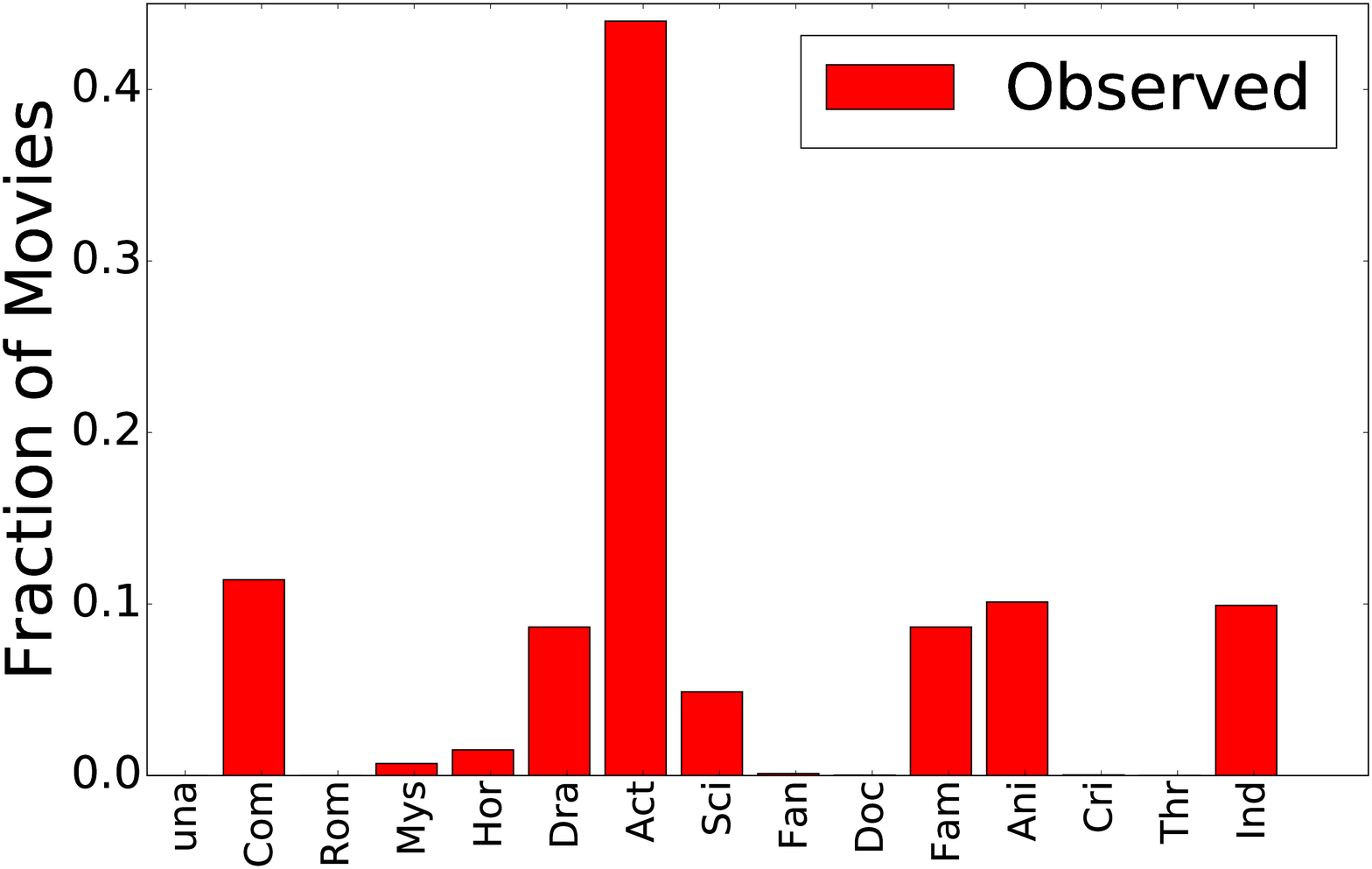}
		\caption{Google Play}
		\label{fig:sub6a}
	\end{subfigure}%
	\hfill
	~\begin{subfigure}{.48\columnwidth}
		\centering
		\includegraphics[width=\textwidth, height=3cm]{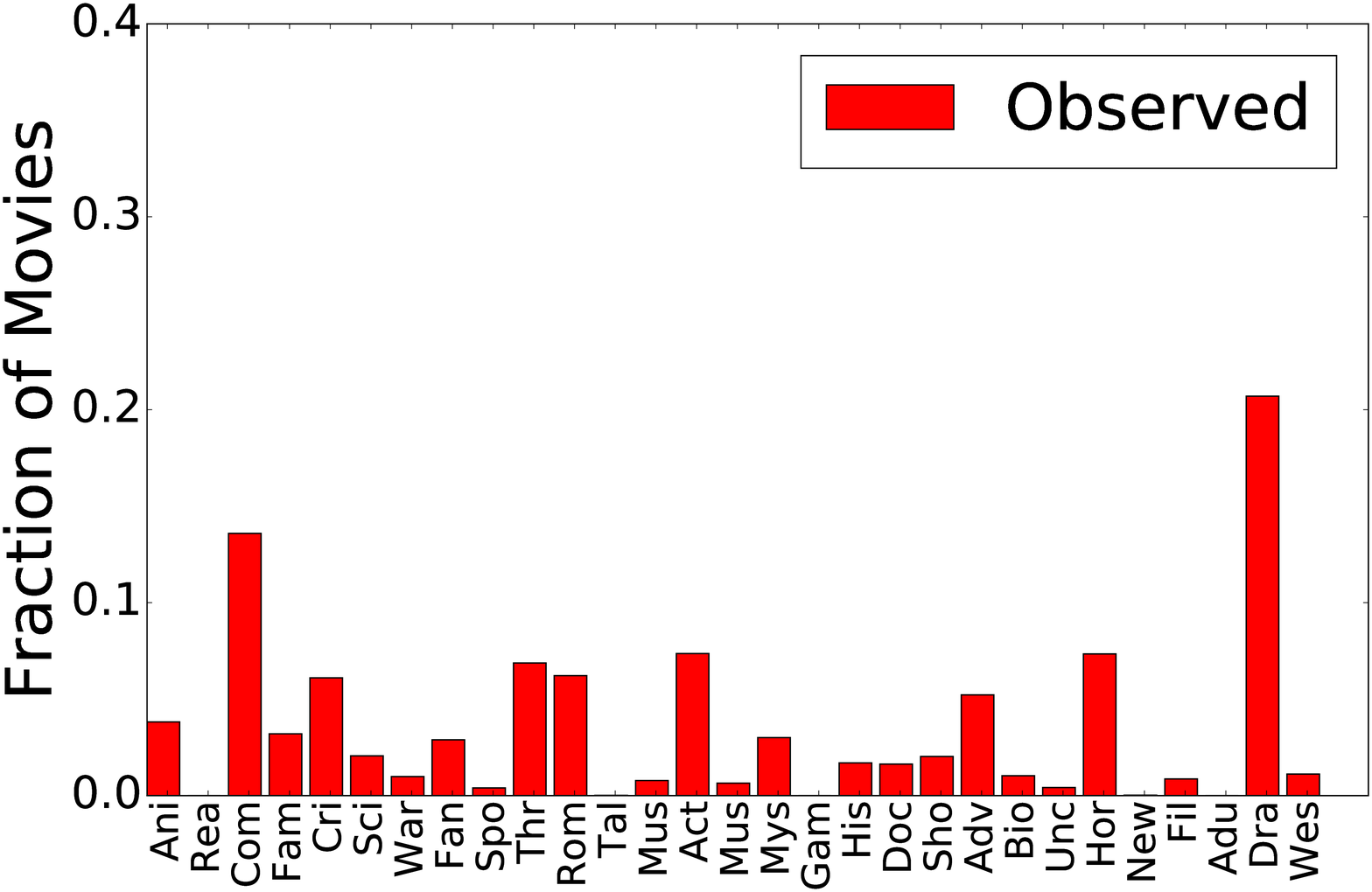}
		\caption{IMDb}
		\label{fig:sub6b}
	\end{subfigure}%
	\caption{{\bf Observed distributions for random walks starting from the movie ``The Godfather'' over different RNs. Length of walk $N = 400$ steps and $t_p = 0.0$ (always following recommendations).}}
	\label{fig:godfather-random-walk-dist}
	\vspace*{-5mm}
\end{figure}

To demonstrate the result of the random walk process described in Section~\ref{sub:Umodel}, Figure~\ref{fig:godfather-random-walk-dist} shows the observed distributions of random walks starting from the movie `The Godfather' over Google Play and IMDb RNs. 
Both walks are for $N = 400$ steps and $t_p = 0.0$ (always following recommendations).
We find that the observed distribution in Google Play is dominated by one genre (`Action', which is one of the genres dominating the product base), while
that in IMDb is much more evenly distributed across genres.

In the rest of this paper, we apply the RN framework on the three aforementioned movie RSs, to measure diversity (Section~\ref{sec:diversity}) and information segregation (Section~\ref{sec:segregation}). 

%% file: Diversity.tex
\section{Diversity in RSs} \label{sec:diversity}

In this section, we apply the proposed framework to measure diversity in various RSs. To begin with, we compute some of the existing diversity measures to show that our framework is very generic. Then, we propose a set of novel diversity metrics that exploits the topology of the RNs.  

\subsection{Existing diversity measures}

We demonstrate that using the proposed RN, we can easily evaluate all the existing diversity measures which do not rely on user-item interaction information.
The in-degree of a node $i$ in the RN refers to the number of times that item $i$ has been recommended on the pages of other items. 
So, the in-degree can be used as an analogy of popularity in the RSs.
Thus, diversity measures such as long-tail novelty \cite{ricci2011introduction} can be computed based on the inverse of the in-degree of various items (nodes).

Most diversity measures consider some notion of similarity (or difference) between two items, which is usually a domain-dependent measure.
For movies, we use the information of the genre of movies to quantify the notion of similarity. As stated in Section~\ref{sub:datasets}, each movie belongs to one or more genres. Let the movie $i$ belong to a set of genres $G(i)$ and movie $j$ belong to the set of genres $G(j)$. We measure the similarity between two movies $i$ and $j$ as
the Jaccard similarity between $G(i)$ and $G(j)$: $sim(i,j) = \frac{|G(i) \cap G(j)|}{|G(i) \cup G(j)|}$. The difference between the two movies is computed as $div(i,j) = 1 - sim(i,j)$. If the edges of the RN are weighted based on the difference measure $div(i,j)$ of the two items, the source-list diversity \cite{bradley2001improving} can be simply evaluated as the average of the weights of the outgoing edges from a node. 
Similarly, the average intra-list diversity \cite{ziegler2005improving} at a certain node $i$ can be measured as the average $div(i,j)$ over all pairs of the out-neighbors of $i$ (to which $i$ links to in the RN). 

Table~\ref{tab:existing-diversity-measures} states the values of some existing diversity measures for the three RSs, as computed using the RN framework. These values stated are the average of the corresponding values for all nodes in the RN.
The Netflix recommendation system is found to have the highest source-list diversity, however IMDb has the highest diversity according to the other three measures.

\begin{table}[tb]
	\small
	\centering
	\begin{tabular}{ |p{3.7cm}|p{1cm}|p{1cm}|p{1.5cm}| }
		\hline
		Diversity measures & \multicolumn{3}{c|}{Recommendation Systems} \\
		\hline
		& IMDb & Netflix & GooglePlay \\
		\hline
		Intra-list diversity \cite{ziegler2005improving} & \textbf{0.6377} & 0.4608 & 0.4205 \\ 
		\hline
		Long-tail novelty \cite{ricci2011introduction} & \textbf{4.5300} & 2.7146 & 3.8872 \\ 
		\hline
		Average unexpectedness\cite{murakami2007metrics} & \textbf{11.6794} & 8.7143 & 7.4879 \\ 
		\hline
		Source-list diversity \cite{bradley2001improving} & 0.6117 & \textbf{0.6988} & 0.4820\\
		\hline
	\end{tabular}
	\caption{{\bf Computing existing diversity measures using the RN framework. IMDb is more diverse according to most measures; Netflix has higher source-list diversity.}}
	\label{tab:existing-diversity-measures}
	\vspace{-6mm}
\end{table}

\subsection{Novel diversity measures}\label{NDRN}

We now propose some novel diversity measures based on the RN framework. 

\noindent \textbf{(1) Assortativity-based measures:}
The diversity of RSs depends on how similar the items recommended on the page of a source item are to the  source item itself. 
The notion of preference of nodes to link to other similar nodes is measured in the complex network literature by the metric {\it assortativity coefficient} (see~\cite{newman2003structure} for definition and mathematical details). 
The assortativity coefficient varies in the range $[-1, 1]$. Negative values suggest that dissimilar nodes are mostly linked; hence, the RSs has higher diversity (and lower similarity of recommendations to the source item). 
On the other hand, positive values indicate that similar nodes are mostly linked (the network is assortative)~\cite{newman2003structure}, hence the diversity of the network is low (though relevance is high).


\if{0}
\begin{table}
	\noindent
	\centering
	\begin{tabular}{l|ccc|c}
		& $t_1$ & $t_2$ & $t_3$ & Sum\\ \hline
		$t_1$ & $e_{11}$ & $e_{12}$  & $e_{13}$ & $a_1$ \\ 
		$t_2$ & $e_{21}$ & $e_{22}$ & $e_{23}$ &  $a_2$ \\
		$t_3$ & $e_{31}$ & $e_{32}$ & $e_{33}$ & $a_3$ \\\hline
		Sum & $b_1$ & $b_2$ & $b_3$ & $n$\\
	\end{tabular}
	\caption{{\bf A sample contingency table for the computation of assortativity coefficient of a network}}
	\vspace*{-5mm}
	\label{AM}
\end{table} 

Let us assume that we can divide the nodes in a network into a few types or classes according to some attribute. For instance, consider a toy example of a network whose nodes can be divided into three types $t_1$, $t_2$, and $t_3$. We count the fraction of edges that link a node of one type to a node of another type. We create a matrix called the `contingency matrix' whose elements are these fractions. 
For instance, Table~\ref{AM} shows a contingency matrix for our toy example, where the element $e_{ij}$ denotes the fraction of edges that link a node of type $t_i$ with a node of type $t_j$. Then assortativity coefficient of the network is defined as follows:
\begin{equation}
	\nonumber
	r = \frac{\sum_i{e_{ii}-\sum_i{a_ib_i}}}{1-\sum_i{a_ib_i}} 
\end{equation}
The assortativity coefficient varies in the range $[-1, 1]$. Negative values of the coefficient suggests that dissimilar nodes are mostly linked in the network. In other words, the RSs has higher diversity (and lower relevance or similarity of recommendations to the source item). 
On the other hand, positive values indicate that similar nodes are mostly linked (the network is assortative)~\cite{newman2003structure}, hence the diversity of the network is low (though relevance is high).

Note that, while the assortativity coefficient gives only a macroscopic measure of the diversity, one can perform more fine-grained analysis on the contingency matrix (Table~\ref{AM}).
For instance, one can investigate questions such as what fraction of the recommendation links from items of one type leads to items of another type. \fi

\if{0}
For the purpose of computing assortativity, various attributes can be used to group the nodes into different types or bins. For instance, if some semantic attribute is known for the items (e.g., genre for movies, or topic for news articles), the items can be binned according to this attribute.
An alternative is `popularity-based binning', where some measure of importance/popularity of the items is used to group the items. For instance, movies can be binned based on their average user-ratings. 
Standard network centrality measures such as in-degree and PageRank~\cite{page1999pagerank} of the nodes can also be used to group the nodes into bins such as `very popular', `moderately popular', and `not popular'. 
Popularity based binning is especially important since a primary motivation of RSs is to push users toward the `long tail', i.e., toward items that are not so popular. Popularity-based binning is an interesting way to measure the extent to which this motivation is getting fulfilled.
\fi

To apply the assortativity-based measures, we 
consider the following attributes for binning the movies.

\noindent \underline{(i) Genre:} It is a natural choice to bin the movies according to their genres. As stated in the dataset section, Google Play specifies $15$ different genres for movies, while IMDb and Netflix specify $29$ different genres.

\noindent  \underline{(ii) Popularity:} 
A natural choice for measuring the popularity of movies would be the ratings or number of views; however, not all movie recommendation sites provide these statistics. Hence, we use network centrality measures as estimates of the popularity of a node (movie). 
To this end, we consider two centrality measures - (1)~in-degree, and (2)~PageRank. We compute both the centrality scores over the RN and normalize to the range $[0.0, 1.0]$; this normalization is needed to compare the values across networks of very different sizes. 
We group the movies into three bins -- (i)~{\it bottom bin}, containing {\it non-popular} nodes (movies) whose centrality is in the range $[0.0, 0.2]$, 
(ii)~{\it middle bin}, containing {\it moderately popular} nodes whose centrality lies in $(0.2, 0.4]$, and (iii)~{\it top bin}, containing {\it very popular} nodes (movies) having centrality higher than $0.4$.

\begin{table}[tb]
	\small
	\centering
	\begin{tabular}{ |p{2.25cm}|p{1cm}|p{1.3cm}|p{1cm}| }
		\hline
		Binning attribute & IMDb & Google Play & Netflix  \\
		\hline
		Genre & 0.1453 & 0.3983 &  0.0729\\
		\hline
		In-degree & 0.1418 & 0.1687 & 0.0314 \\
		\hline
		PageRank & 0.1575 & 0.2525 & 0.0407 \\
		\hline
	\end{tabular}	
	\caption{{\bf Assortativity coefficients of the Recommendation Networks. The lower the assortativity coefficient, the higher is the overall diversity.}}
	\label{tab:assortativity-coeff}
	\vspace*{-7mm}
\end{table}

\vspace{1mm}
\noindent \underline{Computing assortativity coefficients:}
Table~\ref{tab:assortativity-coeff} notes the assortativity coefficients of the three RNs, when the nodes are binned based on the attributes mentioned above.
All RSs have positive coefficients, which is expected because recommendations should be relevant to the source items.
The Netflix recommender system has the lowest assortativity coefficient, indicating highest overall diversity, followed by IMDb.


\begin{figure*}[tb]
	\centering
	\begin{subfigure}{.32\textwidth}
		\centering
		\includegraphics[width=\textwidth, height=4cm]{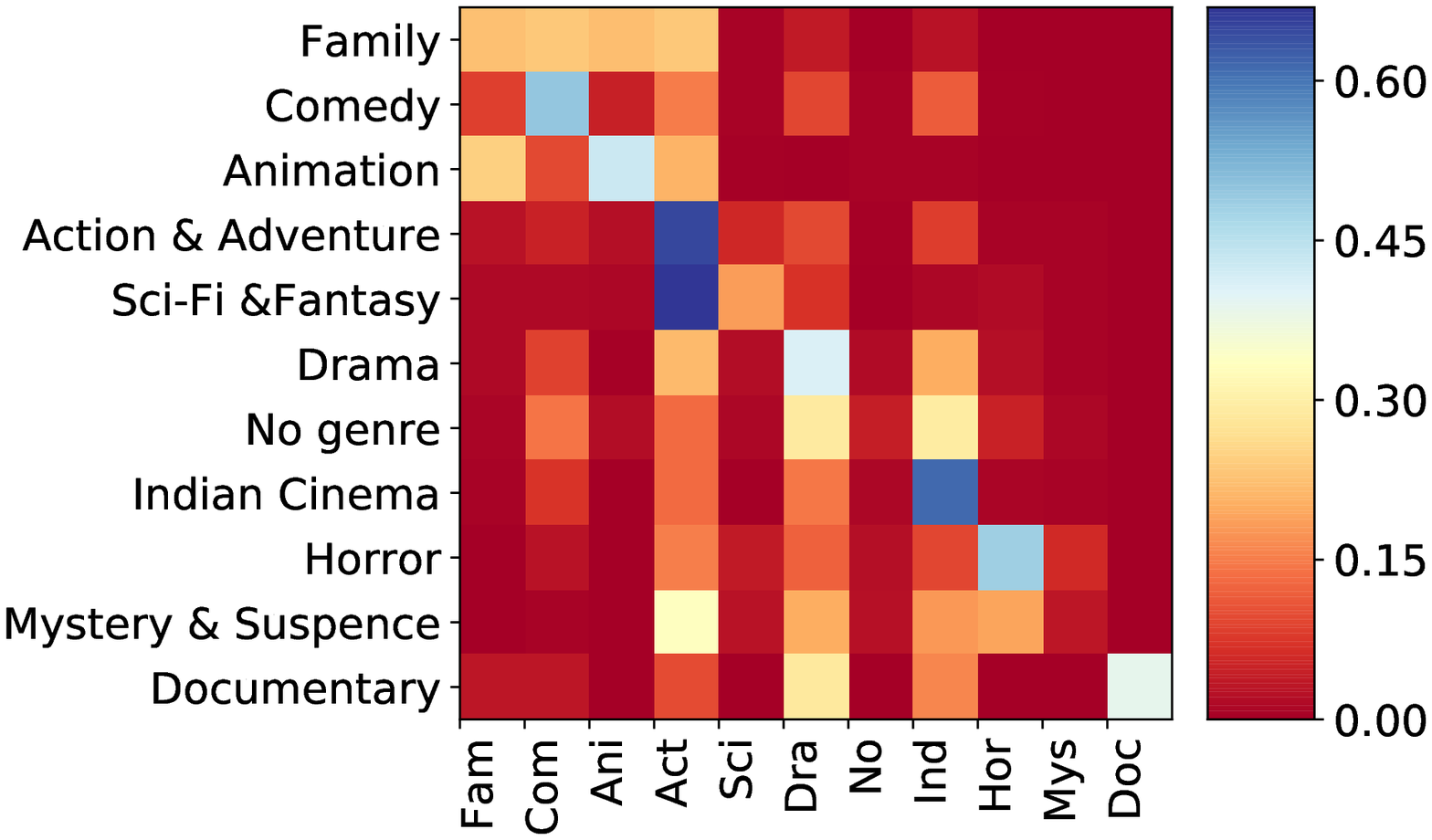}
        \vspace*{-3.5mm}
		\caption{Google Play}
		\label{fig:contingency-matrix-genre-sub1}
	\end{subfigure}%
	\hfill
	~\begin{subfigure}{.33\textwidth}
		\centering
		\includegraphics[width=\textwidth, height=4cm]{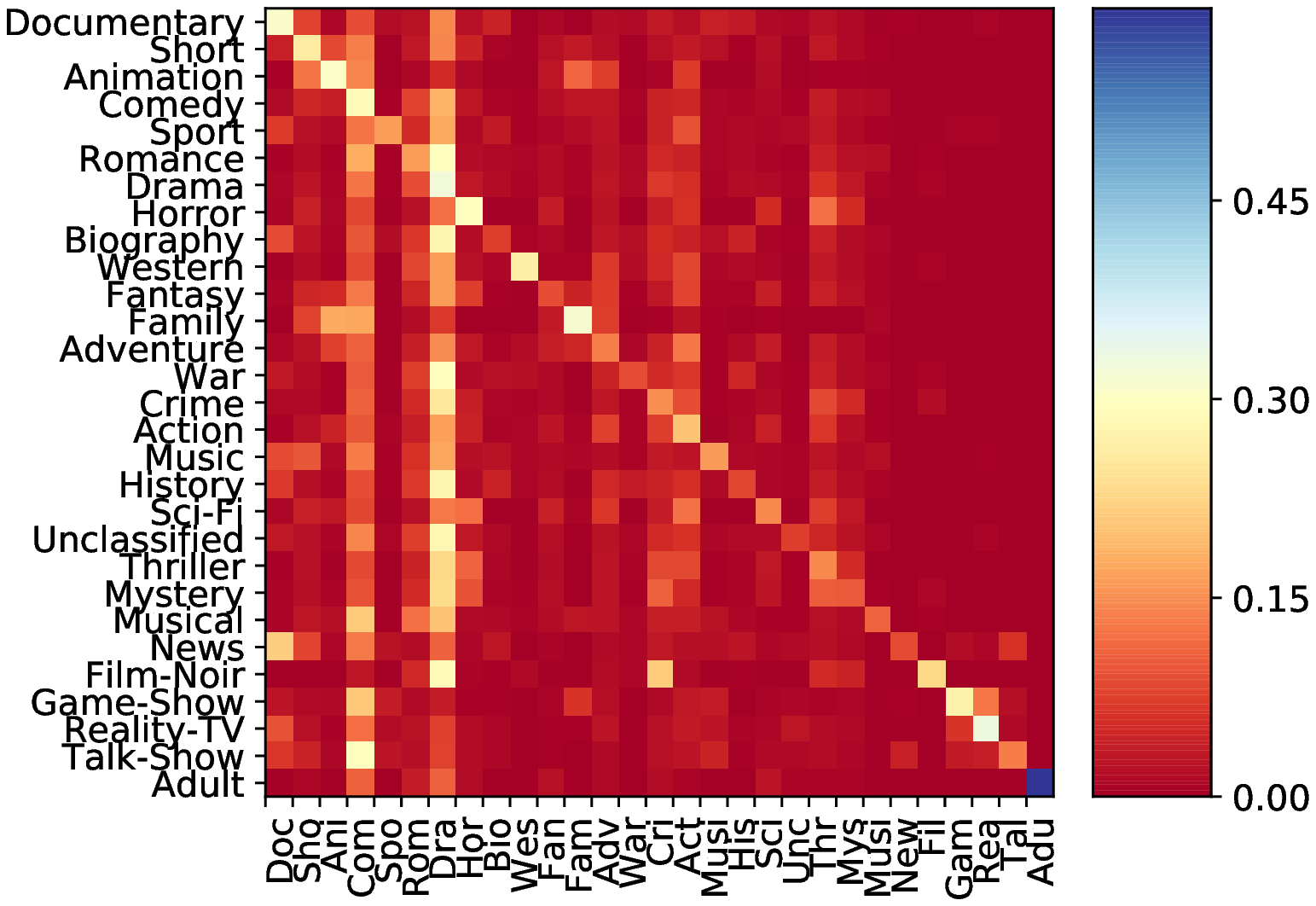}
        \vspace*{-3.5mm}
		\caption{IMDb}
		\label{fig:contingency-matrix-genre-sub2}
	\end{subfigure}
	\hfill
	\begin{subfigure}{.32\textwidth}
		\centering
		\includegraphics[width=\textwidth, height=4cm]{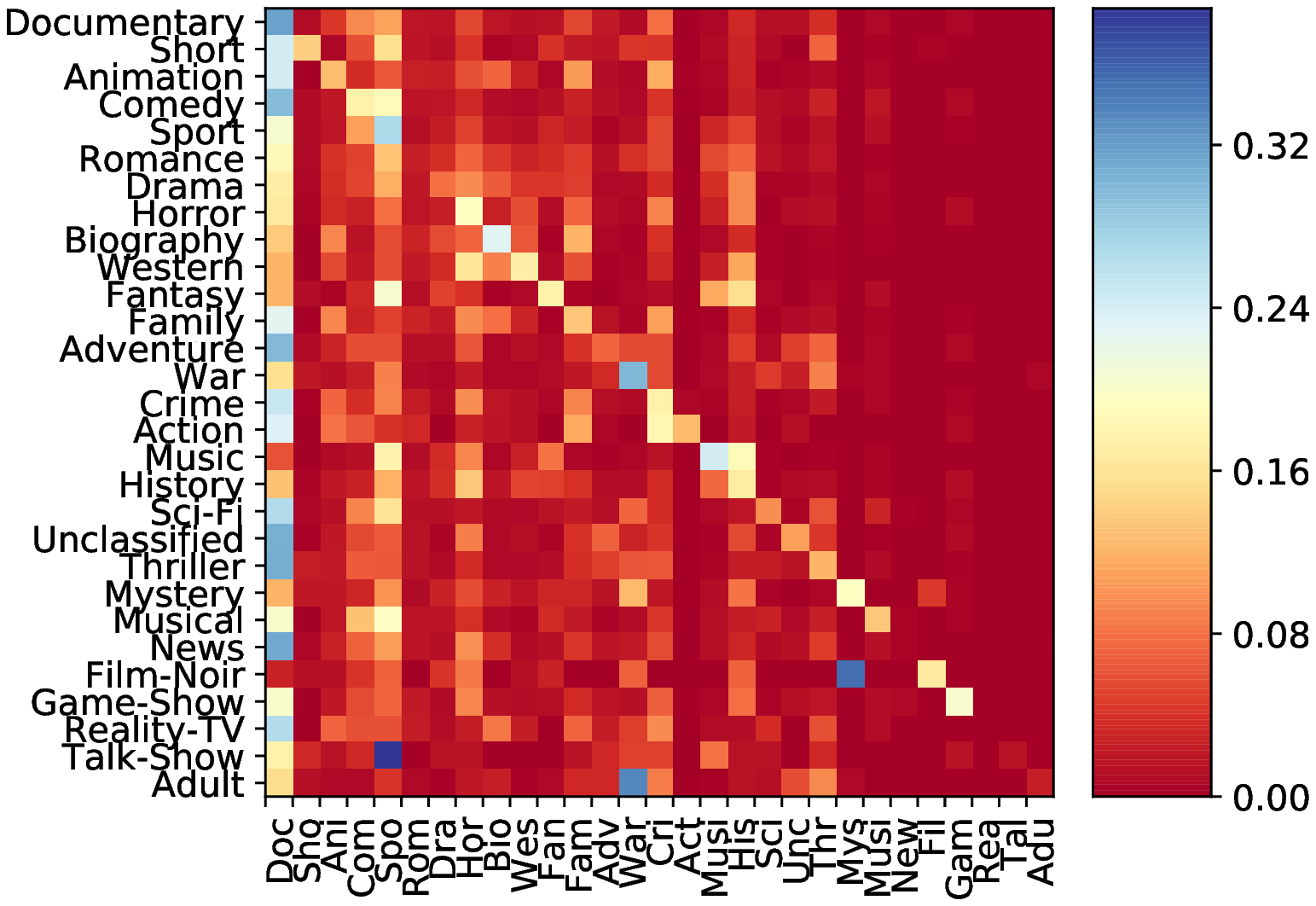}
        \vspace*{-3.5mm}
		\caption{Netflix}
		\label{fig:contingency-matrix-genre-sub3}
	\end{subfigure}
	\vspace*{-2mm}
	\caption{{\bf Visualization of the contingency matrices based on movie genre. The heatmaps show the fraction of outward transitions (recommendations) going from items in one genre to items in another genre.}}
	\label{fig:contingency-matrix-genre}
	\vspace*{-6mm}
\end{figure*}

\noindent \textbf{(2) Measures based on contingency matrix:}
The assortativity coefficient gives only a macroscopic measure of the diversity of a RSs. To perform more fine-grained analysis, we create a matrix called the `contingency matrix’ for a given RN. This matrix is a $m \times m$ matrix, where $m$ is the number of item types (movie genres), and the $(i,j)$-th entry of the matrix denotes the fraction of edges that link a node of type $i$ with a node of type $j$.
We now study the contingency matrix of the three movie RSs. 

\noindent \underline{(i) Contingency matrix based on genres:}
Given the large number of genres (15 for Google Play and 29 for the other two RSs), it is difficult to interpret if numeric entries of the contingency matrix are stated. Hence Figure~\ref{fig:contingency-matrix-genre} visualizes the contingency matrices as heat-maps, where each entry $(i, j)$ measures the fraction of outward edges (recommendations) leading from items in genre $i$, that go to items in genre $j$.

For all the RSs, movies of a particular genre mostly recommend movies of the same genre; this observation is expected, to ensure relevance of the recommendations. Even then, this genre-based analysis reveals that some RSs are more diverse than others. 
For instance, the IMDb RSs (Figure~\ref{fig:contingency-matrix-genre-sub2}) has  higher diversity than the Google Play RSs (Figure~\ref{fig:contingency-matrix-genre-sub1}).  
For Google Play, it can be observed from
Figure~\ref{fig:contingency-matrix-genre-sub1} that movies in $7$ genres - `comedy', `animation', `action and adventure', `Indian cinema', `horror' and `documentary' - almost always recommend movies of the same genre; hence the recommendations lack diversity. 
For IMDb, Figure~\ref{fig:contingency-matrix-genre-sub2} suggests that barring the two genres `adult' and `reality TV', the recommendations from all other genres are quite diverse and lead to many other genres as well. 
The Netflix RSs (Figure~\ref{fig:contingency-matrix-genre-sub3}) also has good diversity.

Also note that movies of few specific genres particularly recommend movies of some other specific genres. 
For instance, in Netflix, `War' movies mostly recommend `adult' movies, while `mystery' movies often recommend `film-noir' movies.
Analyzing the contingency matrices of the RN is a good way of identifying such relationships among genres.

\begin{figure}[tb]
	\begin{subfigure}{\columnwidth}
		\centering
		\includegraphics[width=\textwidth, height=4cm]{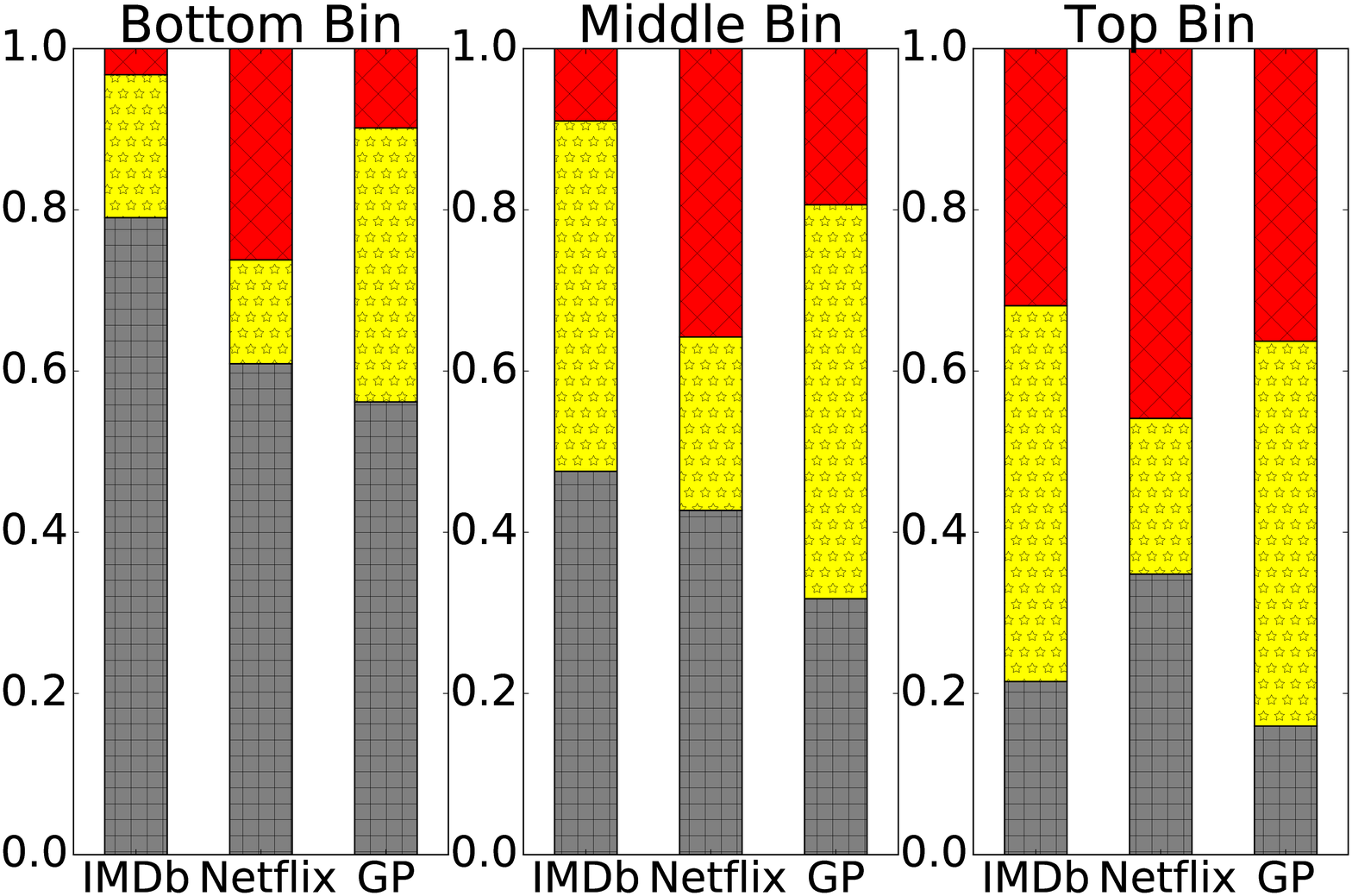}
		\label{fig:sub3}
	\end{subfigure}%
	\vspace*{-5mm}
	\caption{{\bf Visualization of the contingency matrices based on popularity measure: in-degree centrality in the RN. The stacked bars show the normalized fraction of transitions (recommendations) from one bin to another.}}
	
	\label{fig:contingency-matrix-popularity}
	\vspace*{-5mm}
\end{figure}

\noindent \underline{(ii) Contingency matrix based on popularity:}
Binning with respect to in-degree and PageRank centrality measures yielded similar observations for all the RSs; hence we report results for binning based on in-degree. 
Figure~\ref{fig:contingency-matrix-popularity} visualizes the contingency matrices of the RNs based on popularity (in-degree). 
The stacked bars show the fraction of outward transitions (recommendations) going from items in one bin (source bin) to items in another bin (destination bin). 
The grey-colored bars correspond to the recommendations going to the bottom bin (non-popular movies), the yellow bars correspond to recommendations going to the middle bin (moderately popular movies),
and the red bars correspond to recommendations going to the top bin (most popular movies). 
To account for the difference in the number of movies in the different bins, the number of inter-bin transitions has been normalized by the total number of outward edges from nodes in the source bin.

For all the RSs, the top bins have more diverse recommendations, often leading to the middle and bottom bins.
Whereas, the items in the bottom bin (non-popular items in the long tail) mostly recommend other items in the bottom bin itself. 
These observations are expected, because of the motivation of RSs to push users towards the `long tail' (the bottom bin).

Comparing the three RSs, we see that the top bin and bottom bin of IMDb are more diverse than those for Google Play and Netflix. Even though the Netflix top bin recommends a higher fraction of bottom bin movies than in IMDb, the number of recommendations leading to the middle bin is very less throughout. 
In case of Google Play, very few recommendations from the top bin lead to the bottom bin, as compared to the other two RSs. 

\noindent \textbf{(3) Random walk-based measures:}
Most existing diversity measures, including the ones described above, give only a static view of the diversity of a RSs. 
None of these measures can quantify the experience of a user who follows the recommendations of RSs over a period of time.\footnote{Measures which assume availability of user-item interaction information can be used, but as stated earlier, it is practically very difficult to get such information for third party auditors.}
We model such a user $u$ as a random walker over the RN, as described in Section~\ref{sub:Umodel}.   
We consider the binning of the items according to some semantic attribute (e.g., movie genre) or popularity-based attributes, and compute the `observed distribution' $d_u$ of $u$ in terms of the bins. 
We compute the {\it entropy} $H(d_u)$ of this observed distribution to quantify the diversity observed by the user $u$.
Suppose there are $k$ bins or types of the items, $t_1, t_2, \ldots t_k$. User $u$ views $N$ items (i.e., length of the walk is $N$), out of which he views $n_i$ items of the type $t_i$, $i = 1 \ldots k$. Then the entropy is
$H(d_u) = - \sum_{i=1}^{k} \frac{n_i}{N} \cdot \log \frac{n_i}{N}$

Intuitively, the diversity observed by a user depends on the following factors --
(i)~the teleportation probability $t_p$,
(ii)~the length of the walk $N$, and 
(iii)~the first item viewed by the user, i.e., the starting point of the walk.
Note that, in theory, if the random walk is allowed to continue till $d_u$ is stationary, then the observed distribution will be independent of the starting point or the length of the walk. 
However, in practice, a user will only perform a finite walk (view a finite number of movies); hence we consider all the above three factors while quantifying the experience of a user.

For a particular value of $t_p$ and $N$, we start {\it a random walk from every node in the RN}, and compute the average entropy of all the observed distributions (over all walks), so that the {\it final results are independent of any particular starting point}.
We now analyze how average entropy varies with $t_p$ and $N$.

\begin{figure}[tb]
	\centering
	\begin{subfigure}{.25\textwidth}
		\centering
		\includegraphics[width= 4.8cm, height=4cm]{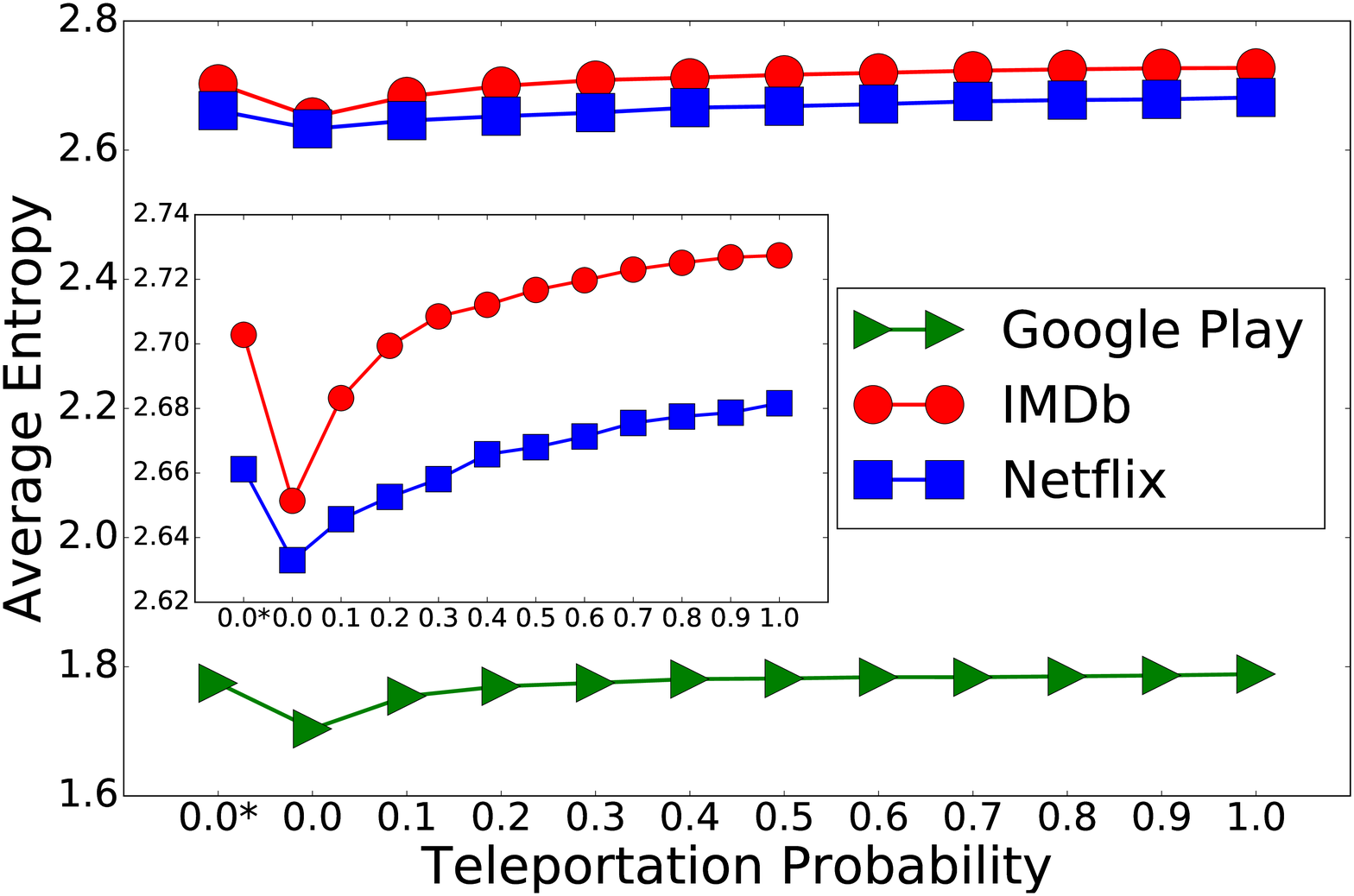}
        \vspace*{-3.5mm}
		\caption{Variation with $t_p$}
		\label{fig:sub7e}
	\end{subfigure}%
	~\begin{subfigure}{.25\textwidth}
		\centering
		\includegraphics[width= 4.8cm, height=4cm]{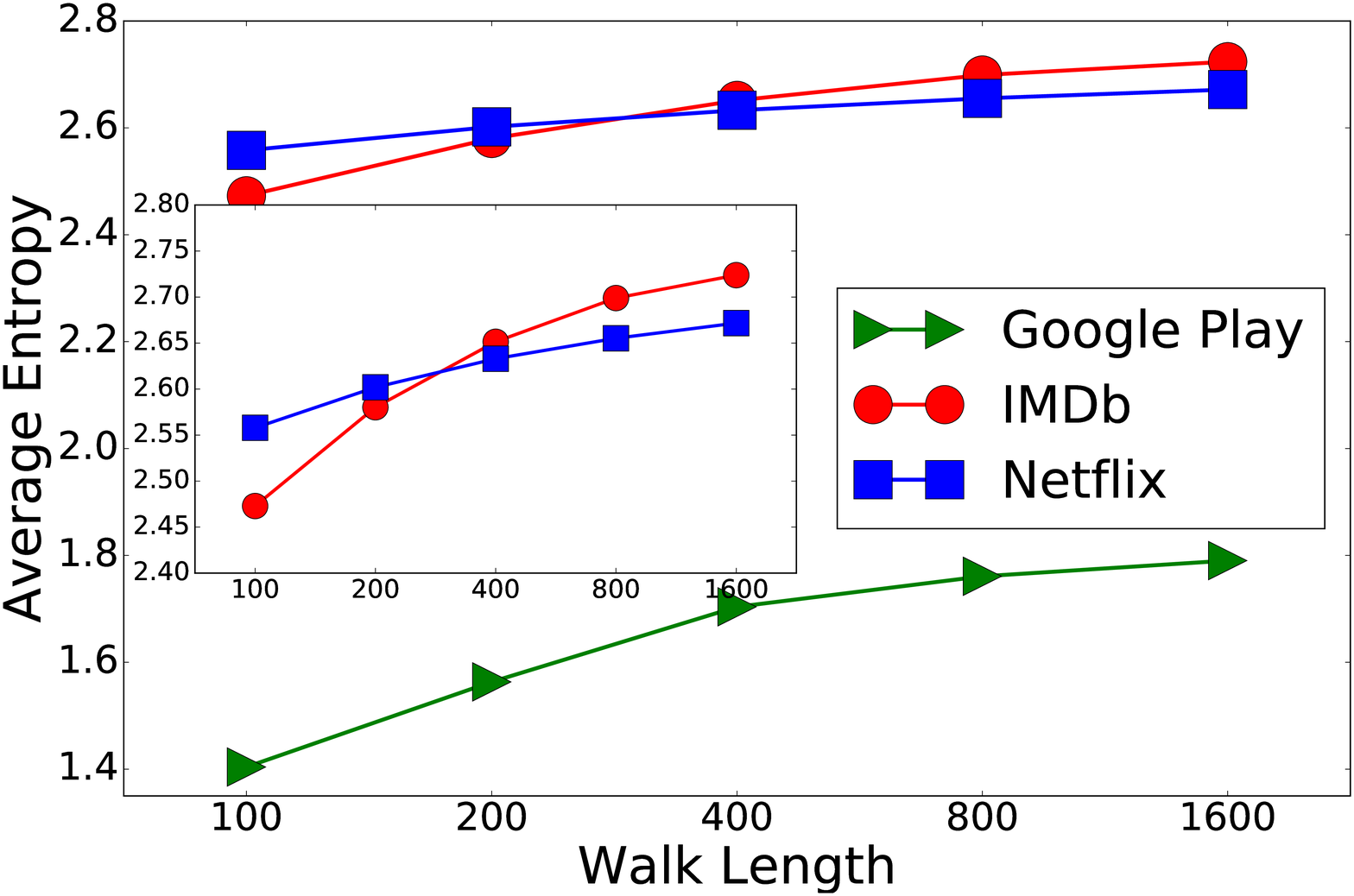}
        \vspace*{-3.5mm}
		\caption{Variation with $N$}
		\label{fig:sub7f}
	\end{subfigure}
    \vspace*{-2mm}
	\caption{{\bf Variation in entropy of observed distribution, with teleportation probability $t_p$ and walk length $N$.}}
	\label{fig:entropy-variation}
	\vspace*{-5mm}
\end{figure}

Figure~\ref{fig:entropy-variation}(a) shows the variation of average entropy with $t_p$ for all the three RSs, keeping $N$ constant at $N=400$. Here the first point in each curve corresponds to the non-stochastic surfer who always chooses the top recommended movie (marked as $t_p = 0.0^{*}$).
Since the curves for IMDb and Netflix are very close together, we magnify these two curves in the inset figure.
The entropy value increases slightly as $t_p$ increases, i.e., as the propensity of a user to follow the recommendations decreases. 
Also, the non-stochastic surfer (a user who always selects the top recommendation) observes higher diversity than the stochastic surfer with $t_p = 0.0$ (who chooses one of the recommendations at random).

Figure~\ref{fig:entropy-variation}(b) shows the variation of average entropy with $N$ for all  three RSs, keeping $t_p = 0.0$ (always choosing one of the recommendations).
The entropy increases with the increase in walk length for all the RSs. However, till a walk length of 200, Netflix shows higher diversity than IMDb, but after that IMDb surpasses Netflix as the walk length increases. 

From both Figure~\ref{fig:entropy-variation}(a) and Figure~\ref{fig:entropy-variation}(b), it is clear that the Google Play RSs has significantly less diversity (entropy), as compared to IMDb and Netflix.

%% file: Information_Segregation.tex
\section{Information Segregation in RSs}  \label{sec:segregation}

An important component of auditing RSs is to determine whether (or to what extent) RSs can lead to information segregation among the users, by exposing different groups of users to different information~\cite{dandekar2013biased,pariser2011filter}.
In this section, we show how the proposed framework can be used to quantify information segregation in RSs.

\subsection{Measures for information segregation}

Following the work of~\cite{dandekar2013biased} we visualize information segregation as a property of the {\it process} which forms the opinion of the society, instead of being a property of the society itself. An individual person's opinion is based on the information that he/she is exposed to throughout the entire process, i.e., throughout his / her interaction with the RSs. 
Hence different groups of users can develop widely varying opinion if they are exposed to different information.


To quantify such phenomena, several information segregation measures were introduced in~\cite{chakraborty2017quantifying}, which follow the rich history of works on residential segregation in a geographical region (such as whether different racial groups are evenly distributed in a city)~\cite{dandekar2013biased}.
To apply these measures, the information content in a system is modeled as a $m$-dimensional Euclidean space, where $m$ is the total number of different information units.
We refer to this $m$-dimensional space as the {\bf information space}.
In our context of movie RSs, the {\bf information units} are the different genres of movies,
and each individual movie is an {\bf information source}.
As stated in section~\ref{sub:datasets}, a movie can have multiple genres; hence, a particular information source can be mapped to multiple information units. Genre of movies could be thought of as analogous to the different types of news in news media domain. In this section, therefore, we primarily focus on the analysis in an information-space where each dimension is essentially a genre.

Out of the different measures introduced in~\cite{chakraborty2017quantifying}, we consider the following two measures that 
attempt to capture the notion of whether different groups of users in the society are being exposed to different information units by the RSs. 


\noindent \textbf{(1) Evenness} is a measure that captures how uniformly members of a particular group are exposed to different information units in the information space.
    For a particular group of users $A$, the Gini coefficient $G_A$ measures the {\it un-evenness} within the group, by capturing the mean absolute difference between the visibility of different information units across the members of the group. 
    Subtracting $G_A$ from $1$ gives us the information evenness $IE_A$: $IE_A = 1 - G_A = 1- \frac{\sum_{i=1}^{m}{\sum_{j=1}^{m}{|a_i-a_j|}}}{2*m*a_{total}}$ where, $m$ is the total number of information units (movie genres), $a_i$ is the total number of information sources (movies) of the information unit $i$ that have been seen by the users of group $A$, $a_{total}$ is the total number of information sources seen by users in group $A$. 
        
This measure essentially captures the breadth of the exposure that a group of users has while interacting with the RSs over a period of time. $IE_A$ varies between $0.0$ and $1.0$; higher evenness indicates that the group of users has exposure to information of most of the $m$ different information units, without being segregated or polarized to only a few information units. 
The higher the evenness, the better is the distribution of visibility of different information units, hence the lesser polarizing (segregating) is the system.
        

\vspace{2mm}
    \noindent \textbf{(2) Concentration} of a user-group $A$ refers to the relative fraction of the universe of all items, that $A$ has been exposed to. 
    The information concentration $IC_A$ of group $A$ is defined as: $IC_A =\frac{1}{2}*\sum_{i=1}^{m}{\frac{a_i}{a_{total}} * \frac{n_i}{n_{total}}}$ where $a_i, a_{total}$ and $m$ are as defined earlier, $n_i$ is the number of sources in information unit $i$, and $n_{total}$ is the total number of information sources in the whole system.
        
This measure essentially captures the depth of the exposure that a group of users has over different information units, while interacting with the RSs over a period of time. In our context, this measure captures the number of movies of different genres that have been viewed by members of a group. $IC_A$ varies between $0.0$ and $1.0$; higher the $IC_A$, lesser is the concentration. Lower values of concentration for a given group of users implies that, the users have exposure to information sources (movies) that map to information units (movie genres) spanning a larger part of the information space (universe of all movies), without being segregated or polarized to only a small part of the information space.
        

Note that the concentration measure is fundamentally different from the evenness measure. Different information units may have different number of information sources, with some units having more sources compared to others (in our context, different genres can have different number of movies). Therefore, even though two groups $A$ and $B$ have been exposed to the same number of information units (genres), i.e., their evenness is the same, their concentration can be different because the exposed units can span different proportions of the information space.


\subsection{Using proposed framework to measure information segregation}

As stated earlier, we model the interaction of an individual user with the RSs as a random walk over the RN. We divide a population of users into several groups, based on their starting movie (node) and propensity to follow the recommendations (teleportation probability $t_p$). All users in a particular group start from the same movie, and have the same $t_p$.

To model users having many different starting points and $t_p$, we do the following. We select ten movies from the all-time top-50 movies of IMDb~ \cite{Top250} as starting nodes for the random walks.
The reason behind selecting popular movies is that we assume most users are likely to start browsing an online movie site with a popular movie (and will then follow recommendations for further exploration for $N = 400$). 
Also, we select only those movies as starting points that are present in all the RSs under consideration (IMDb, Google Play, Netflix). 
For $t_p$, we consider the values $0.0, 0.1, 0.2, \ldots, 1.0$, i.e., $11$ distinct values. We consider a distinct group of users starting from each of the $10$ starting points, and having each of the $11$ values for $t_p$. Thus, we simulate a total of 110 user-groups. We consider each group to have 10 members, and while reporting observations for a group, we consider the average over all members to remove statistical variations. 

We simulate exactly the same 110 groups performing random walks over all the different RNs (IMDb, Netflix, Google Play). We consider the observed distributions of the different users after the random walks are completed, and identify the extent of information segregation in different RSs by applying the two measures defined above. 
While reporting results for a particular RSs, we average the results across all the starting nodes, so that the final reported results are independent of a particular start point.


\begin{figure}[tb]
	\centering
	\begin{subfigure}{.25\textwidth}
		\centering
		\includegraphics[width= 4.8cm, height=3.5cm]{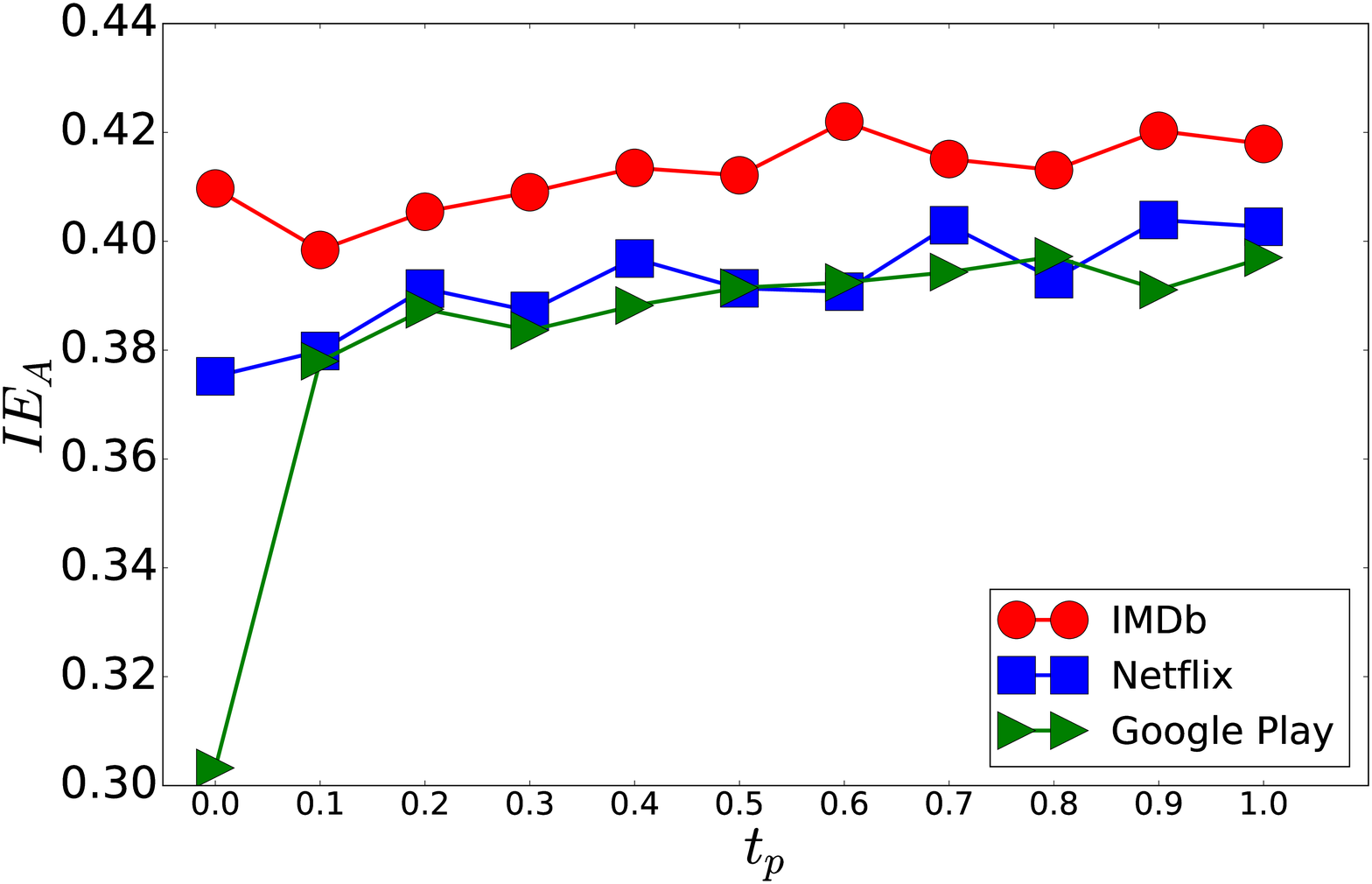}
		\caption{Evenness vs. $t_p$.}
		\label{fig:subga}
	\end{subfigure}%
	\begin{subfigure}{.25\textwidth}
		\centering
		\includegraphics[width= 4.8cm, height=3.5cm]{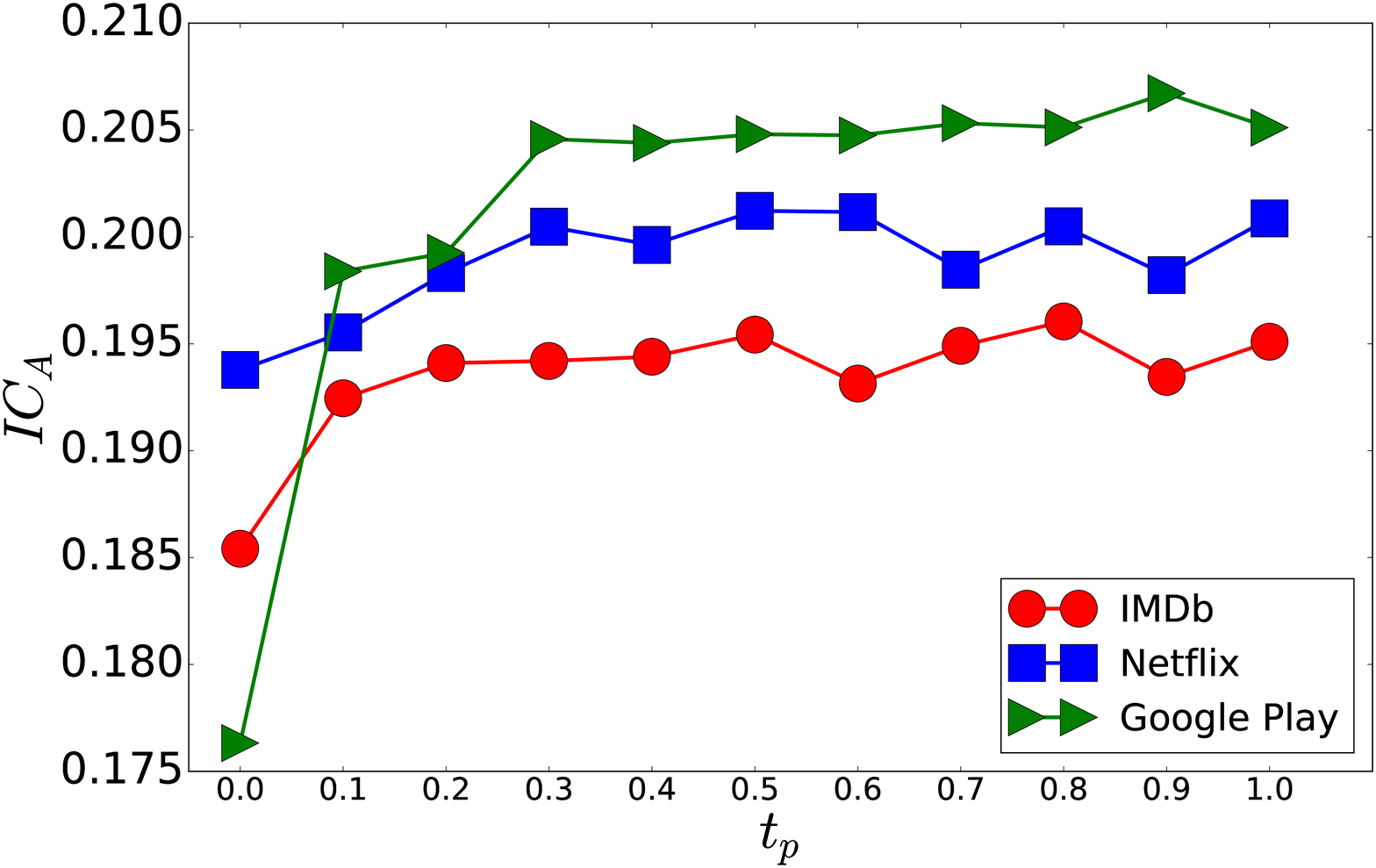}
		\caption{Concentration vs. $t_p$.}
		\label{fig:subgb}
	\end{subfigure}
	\caption{{\bf Variation in evenness and concentration of the observed distributions for different teleportation probabilities. Results are averaged over all groups starting their walks from 10 popular starting points.}}
	\label{fig:KS}
	\vspace*{-5mm}
\end{figure}

\subsection{Information segregation of movie RSs}


Figure~\ref{fig:KS} shows the variation of evenness and concentration with $t_p$ (the propensity of members of a group to follow recommendations). As stated above, the results have been averaged over all groups starting their walks from the 10 popular starting points. 

Figure~\ref{fig:KS}(a) shows that evenness stays more or less stable for IMDb, irrespective of to what extent a user follows the recommendations. However, evenness increases in case of Netflix and Google Play as users become {\it less} likely to follow the recommendations. These observations suggest that, following the IMDb recommendations a user will have a considerably less polarizing / segregating experience of different movie genres, as compared to that for Netflix and Google Play. Interestingly, for Netflix and Google Play, evenness increases with the increase in $t_p$, i.e., a random sampling will give a user less segregated views than following the recommendations. 
Note that IMDb exhibits a different behavior compared to the two service provider sites. Among the two service providers Netflix's evenness is closer to that of IMDb compared to Google Play. 
An important point to note is that in case of Google Play, for $t_p=0$, i.e., following the underlying RN results in 8\% lesser evenness compared to the other values of $t_p$. Overall, the evenness of IMDb is 11\% higher than that in Google Play and nearly 4\% better than the that of Netflix RN.

From Figure~\ref{fig:KS}(b), we find that the concentration measures show very comparable results for all three RSs. Also,
the concentration remains quite stable irrespective of the teleportation probability, for all the three RSs. 
The Google Play RN seems to be less concentrated than both IMDb and Netflix over most of the $t_p$ values (higher $IC_A$ implies lower concentration). Interestingly, at $t_p=0$ the concentration of Google Play is the highest among all the three RN. This implies that the recommendations given by Google Play are considerably more segregated as compared to IMDb and Netflix.
Again it can be seen in Figure~\ref{fig:KS}(a) and (b) that the curves for the two service provider systems Netflix and Google Play are close together, while that for IMDb shows slightly different behavior.



%% file: Conclusion.tex
\section{Concluding Remarks}

We propose a novel network-based framework for auditing RSs, especially with respect to diversity and polarization. 
The framework can not only be used to compute existing diversity measures, but also provides novel diversity measures based on mixing patterns in the recommendation networks. 
Additionally, while most existing measures are static, the proposed framework helps to 
analyze the experience of users who use the RSs over a long period of time.
The proposed framework is suitable for use by third-party auditors, since it does not rely on user-item interaction information which is practically never public.

\if{0}
In the present work, we used the framework to audit and compare three popular movie RSs. 
We demonstrated how the framework can be used to analyze properties of the RSs such as
how movies of one genre are recommending movies of other genres (Figure~\ref{fig:contingency-matrix-genre}), to what extent the RSs is pushing users to the newer and less popular items (Figure~\ref{fig:contingency-matrix-popularity}), and so on.
Comparing the three movie RSs, we gain some interesting insights: \textbf{(a)} The IMDb RSs is significantly more diverse than Netflix and Google Play, as is evident from the assortativity coefficients in Table~\ref{tab:assortativity-coeff}. The information segregation analysis in terms of evenness (Figure~\ref{fig:subga}) further emphasizes the higher diversity of IMDb. \textbf{(b)} In contrast, the Google Play recommendations are found to be considerably more segregating and less diverse, as shown by the high assortativity values in Table~\ref{tab:assortativity-coeff}, as well as by the analysis of the concentration measure of information segregation. \textbf{(c)} Analyzing the popularity of source and recommended items (Figure~\ref{fig:contingency-matrix-popularity}) suggests that IMDb recommendations push the users more smoothly toward the long-tail (the unpopular items), as compared to the other two RSs. 
Especially, Google Play is very assortative, giving very few recommendations to the unpopular or moderately popular movies.  \textbf{(d)} Irrespective of the propensity of a user to follow the recommendations, IMDb always provides recommendations which largely covers the entire information space. However, in case of Netflix and Google Play, the possibility of exploring the entire information space increases for a user as he/she gradually stops following the recommendations (Figure~\ref{fig:subga}).
\fi

\noindent
The various insights obtained in the experiments show that IMDb RN is significantly different from the RN of the service provider platforms. 
For instance, IMDb has the highest diversity; in contrast, the Google Play recommendations are considerably more segregating and less diverse.
Observations such as the above would help third-party auditors (as well as the designers of the RSs) to gain important insights about the functioning of the RSs.
Furthermore, apart from auditing existing RSs, the RN framework can also be extended to indicate potential ways to improve RSs, e.g., by re-wiring the RNs. We plan to explore this direction in future.